\RequirePackage{ifpdf}
\ifpdf
\else
\PassOptionsToPackage{dvipdfmx}{graphicx}
\PassOptionsToPackage{dvipdfmx}{hyperref}
\fi

\documentclass[12pt,a4paper]{article}
\usepackage{jheppub}
\usepackage[utf8]{inputenc}

\allowdisplaybreaks[1]
\usepackage{amsmath,bm}
\usepackage{subfigure}
\usepackage{amssymb}
\usepackage{colortbl}

\title{
Multioscillating black holes
}

\author[1]{Takaaki~Ishii,}
\author[2]{Keiju~Murata,}
\author[3]{Jorge~E.~Santos,}
\author[4]{Benson~Way}

\affiliation[1]{Department of Physics, Kyoto University, Kitashirakawa Oiwake-cho, Kyoto 606-8502, Japan}
\affiliation[2]{Department of Physics, College of Humanities and Sciences, Nihon University, Sakurajosui, Tokyo 156-8550, Japan}
\affiliation[3]{DAMTP, Centre for Mathematical Sciences, University of Cambridge, Wilberforce Road, Cambridge CB3 0WA, UK}
\affiliation[4]{Departament de F\'{i}sica Qu\`{a}ntica i Astrof\'{i}sica, Institut de Ci\`{e}ncies del Cosmos\\
Universitat de Barcelona, Mart\'{i} i Franqu\`{e}s, 1, E-08028 Barcelona, Spain}
\emailAdd{ishiitk@gauge.scphys.kyoto-u.ac.jp}
\emailAdd{murata.keiju@nihon-u.ac.jp}
\emailAdd{jss55@cam.ac.uk}
\emailAdd{benson@icc.ub.edu}

\abstract{%
We study rotating global AdS solutions in five-dimensional Einstein gravity coupled to a multiplet complex scalar within a cohomogeneity-1 ansatz. The onset of the gravitational and scalar field superradiant instabilities of the Myers-Perry-AdS black hole mark bifurcation points to black resonators and hairy Myers-Perry-AdS black holes, respectively. These solutions are subject to the other (gravitational or scalar) instability, and result in hairy black resonators which contain both gravitational and scalar hair. The hairy black resonators have smooth zero-horizon limits that we call graviboson stars. In the hairy black resonator and graviboson solutions, multiple scalar components with different frequencies are excited, and hence these are multioscillating solutions. The phase structure of the solutions are examined in the microcanonical ensemble, i.e. at fixed energy and angular momenta. It is found that the entropy of the hairy black resonator is never the largest among them.  We also find that hairy black holes with higher scalar wavenumbers are entropically dominant and occupy more of phase space than those of lower wavenumbers.
}

\preprint{KUNS-2856}

\begin{document}
\maketitle

\section{Introduction}
\label{sec:intro}
Through superradiant scattering, energy can be extracted from rapidly rotating black holes (see \cite{Brito:2015oca} for a review).  In global anti-de Sitter space (AdS), the reflecting boundary causes these black holes to be unstable to the superradiant instability \cite{Hawking:1999dp,Reall:2002bh,Cardoso:2004hs,Kunduri:2006qa,Cardoso:2006wa,Murata:2008xr,Kodama:2009rq,Dias:2011at,Dias:2013sdc,Cardoso:2013pza}, whose ultimate endpoint remains an open problem \cite{Dias:2011ss,Dias:2015rxy,Niehoff:2015oga,Chesler:2018txn}.  Because small black holes typically have high angular frequency, the ultimate configuration of low-energy states in AdS likewise remains unknown. 

Early work on this problem involved studying perturbations of the Kerr-AdS and Myers-Perry-AdS black holes 
(the higher-dimensional analog of Kerr-AdS \cite{Myers:1986un,Hawking:1998kw,Gibbons:2004uw,Gibbons:2004js}, see \cite{Emparan:2008eg} for a review), 
where quasi-normal spectra were obtained in \cite{Dias:2013sdc,Cardoso:2013pza}.
For specific modes, the onset of the superradiant instability occurs at a frequency where $\Im(\omega)=0$, but $\Re(\omega)\neq0$.   This suggests the existence of a time-periodic black hole with a single helical Killing vector that branches from these onsets.  Such black holes, called \emph{black resonators}, which can be viewed as black holes with a gravitational hair, were later constructed in \cite{Dias:2015rxy}, where it was found that they have higher entropy (horizon area) than the corresponding Kerr-AdS black hole with the same mass and angular momentum.

It is therefore entropically permissible for Kerr-AdS black holes to evolve towards black resonators.  However, though black resonators are stable to the mode that generated them, they are still rapidly rotating and hence remain unstable to other, typically higher, superradiant modes \cite{Dias:2011ss,Dias:2015rxy,Niehoff:2015oga,Chesler:2018txn,Green:2015kur}.  It appears, therefore, that the instability leads to a cascade with higher and higher modes growing in time.  If there is indeed an unceasing energy cascade towards higher modes, there will eventually be a significant amount of energy placed in sub-Planckian length scales, which can be viewed as a violation of the weak cosmic censorship conjecture \cite{Penrose:1969pc}.

To date, there is only a single study of time-evolution involving the rotational superradiance of AdS \cite{Chesler:2018txn}.  In Kerr-AdS, there is typically one unstable mode that dominates the dynamics at early times.  Evolution then proceeds towards a black resonator until the instabilities of the black resonator itself begin to take over and drive the continuing evolution.  The evolution in \cite{Chesler:2018txn} was not continued further due to numerical limitations.

Because of the lack of symmetries and the long time-scales involved, the study of the superradiant instability is a significant numerical challenge.  It is therefore fortunate that a simplification in a more limited setting has been found. By moving to five dimensions and allowing both angular momenta to be equal, black resonators with a cohomogeneity-1 ansatz (i.e. the metric functions depend only on a single variable, and the solution can be obtained by solving ODEs) were constructed in \cite{Ishii:2018oms}. The scalar, electromagnetic, and gravitational quasinormal modes of these black resonators were studied shortly thereafter in \cite{Ishii:2020muv}.\footnote{For the superradiant instability of an electromagnetic perturbation isolated from these modes, a cohomogeneity-1 photonic black resonator was also constructed in \cite{Ishii:2019wfs}.}  As anticipated, black resonators are unstable to higher modes.

These higher-mode instabilities of black resonators also have their individual onsets, from which new black resonators having multiple frequencies could be generated.  
We will refer such solutions as multi black resonators.
Here, we set out to construct such multi black resonator solutions and to study their relationship with black resonators and Myers-Perry-AdS black holes. Ideally, we would study multi black resonators that are generated by gravitational perturbations of black resonators.  However, these perturbations break too many symmetries of the original black resonator solution.  We therefore focus on multi black resonators that are generated by scalar fields.  Ordinarily, a scalar field would also break most of these symmetries, but we will rely on a multiplet scalar constructed using Wigner D-matrices, from which the cohomogeneity-1 structure can be preserved.  A scalar doublet version of such a system was previously studied in \cite{Dias:2011at,Choptuik:2017cyd}\footnote{See also \cite{Stotyn:2013yka,Stotyn:2013spa} for a study covering different number of dimensions.}. The higher scalar multiplets we introduce can coexist with oscillations of the metric, but do not contribute additional extra oscillating frequencies to the metric.

Even in this limited setting, the full space of solutions is intricate.  There are Myers-Perry-AdS black holes, black holes with scalar hair, black resonators, and now hairy black resonators, which are a kind of multi black resonator.  In addition, there are boson stars, geons, and now graviboson stars, which are all horizonless solutions that serve as the zero-size limit of hairy Myers-Perry-AdS, black resonators, and hairy black resonators, respectively.  All of these solutions compete thermodynamically when they share the same energy and angular momenta.  We will compute the full phase diagram of this system.  Perhaps surprisingly, we find that hairy black resonators are never dominant in such a phase diagram.

Another advantage of the multiplet scalar model is that different nonlinear solutions generated by different mode-instabilities can be consistently compared with one another, while maintaining a cohomogeneity-1 ansatz.  This will allow us to show that black holes generated from higher scalar mode instabilities have higher entropy and occupy a larger region of phase space than those from lower modes.  Similar conclusions reached by previous work were only argued by extrapolating perturbative calculations, and here we are able to perform full nonlinear calculations and compute the actual phase boundaries.

This paper is structured as follows.  In the next section, we review some basic properties of isometries of $S^3$ and Wigner D-matrices that we use in the construction of our ansatz. Then in section \ref{sec:scalar_coh1}, we describe details of our ansatz.  Sections \ref{sec:geonMPAdS} and \ref{sec:BR} review the Myers-Perry-AdS solution, geons, and black resonators in this ansatz, which were studied in \cite{Ishii:2018oms,Ishii:2020muv}.  Section \ref{sec:HMPAdS} discusses hairy black holes and boson stars which are higher multiplet versions of those in \cite{Dias:2011at}.  Then, in section \ref{sec:HBR}, we present the entirely new hairy black resonator and graviboson star solutions.  The entire phase diagram of all solutions then pieced together in section \ref{sec:phase}, and then we compare the results to a higher wavenumber in section \ref{sec:phasej5}.  We then finish with some concluding remarks in section \ref{sec:conclude}.  The appendix contains technical details.

\section{Hypersphere isometries and Wigner D-matrices}

Our metric ansatz will contain deformations of an $S^3$ whose perturbations can naturally be written in terms of Wigner D-matrices $D^j_{mk}(\theta,\phi,\chi)$.  Let us therefore begin with a review of the Wigner D-matrices, which we will later use for designing a cohomogeneity-1 scalar field ansatz.
(See Refs.~\cite{Sakurai:2011zz,Hu:1974hh,Murata:2007gv,Kimura:2007cr,Murata:2008yx,Murata:2008xr,Ishii:2020muv}
for an introduction to the Wigner D-matrix and its applications in gravitational perturbation theory.)

We focus on the $SO(4)\simeq SU(2)_L \times SU(2)_R$ isometry of $S^3$, whose metric can be written as
\begin{equation}
\mathrm{d}\Omega_3^2=\frac{1}{4}(\sigma_1^2+\sigma_2^2+\sigma_3^2)\ ,
\end{equation}
where $\sigma_i \ (i=1,2,3)$ are 1-forms defined by
\begin{equation}
\begin{split}
\sigma_1 &= -\sin\chi \mathrm{d}\theta + \cos\chi\sin\theta \mathrm{d}\phi\ ,\\
\sigma_2 &= \cos\chi \mathrm{d}\theta + \sin\chi\sin\theta \mathrm{d}\phi\ ,\\
\sigma_3 &= \mathrm{d}\chi + \cos\theta \mathrm{d}\phi  \ .
\end{split}
\label{inv1form}
\end{equation}
These satisfy the $SU(2)$ Maurer-Cartan equation $\mathrm{d}\sigma_i = (1/2) \epsilon_{ijk} \sigma_j \wedge \sigma_k$.
The coordinate ranges are $0\leq \theta < \pi $, $0\leq \phi <2\pi$, and $0\leq \chi <4\pi$, and have a twisted periodicity $(\theta,\phi,\chi) \simeq (\theta,\phi+2\pi,\chi+2\pi) \simeq (\theta,\phi,\chi+4\pi)$.
The Killing vectors generating $SU(2)_L$ and $SU(2)_R$, denoted by $\xi_i$ and $\bar{\xi}_i$, respectively, are given by
\begin{equation}
\begin{split}
\xi_1 &= \cos\phi\partial_\theta +
\frac{\sin\phi}{\sin\theta}\partial_\chi -
\cot\theta\sin\phi\partial_\phi\ ,\\
\xi_2 &= -\sin\phi\partial_\theta +
\frac{\cos\phi}{\sin\theta}\partial_\chi -
\cot\theta\cos\phi\partial_\phi\ ,\\
\xi_3 &= \partial_\phi\ ,
\end{split}
\label{lkv}
\end{equation}
and
\begin{equation}
\begin{split}
\bar{\xi}_1 &= -\sin \chi \partial_\theta + \frac{\cos \chi}{\sin \theta} \partial_\phi - \cot \theta \cos \chi \partial_\chi \ , \\
\bar{\xi}_2 &= \cos \chi \partial_\theta + \frac{\sin \chi}{\sin \theta} \partial_\phi - \cot \theta \sin \chi \partial_\chi \ , \\
\bar{\xi}_3 &= \partial_\chi \ .
\end{split}
\label{rkv}
\end{equation}
Note that $\bar{\xi}_i$ are the dual vectors of $\sigma_i$:
$(\sigma_i)_\alpha (\bar{\xi}_j)^\alpha=\delta_{ij} \ (\alpha=\theta,\phi,\chi)$.

Using language from quantum mechanics, we can define the ``angular momentum'' operators
\begin{equation}
L_i= i\xi_i \ ,\quad R_i= i \bar{\xi}_i \ ,
\label{SU2LR}
\end{equation}
which satisfy the commutation relations $[L_i,L_j]=i \epsilon_{ijk}L_k$ and $[R_i,R_j]=-i \epsilon_{ijk}R_k$.
These operators are Hermitian under the inner product on the $S^3$,
\begin{equation}
 (f,g)\equiv 
\frac{1}{8}\int^\pi_0\mathrm{d}\theta\int^{2\pi}_0\mathrm{d}\chi \int_0^{4\pi}\mathrm{d}\chi \,  \sin\theta  f^\ast(\theta,\phi,\chi)g(\theta,\phi,\chi)\ .
\label{innerprod}
\end{equation}

Under the $SU(2)_L$ and $SU(2)_R$, the 1-forms introduced in Eq.~(\ref{inv1form}) transform as
\begin{equation}
 L_i \sigma_j=0\ ,\quad R_i \sigma_j = -i\epsilon_{ijk}\sigma_k\ ,
\label{LRsigma}
\end{equation}
where the operations of $L_i$ and $R_i$ are defined by Lie derivatives.
From the first equation of \eqref{LRsigma}, one can see that $\sigma_i$ are invariant under $SU(2)_L$. For this reason, they are called $SU(2)$-invariant 1-forms.
The second equation means that $R_i$ generate the three-dimensional rotation of the ``vector'' $(\sigma_1,\sigma_2,\sigma_3)$.
In particular, $R_3$ generates $U(1)_R \subset SU(2)_R$, which corresponds to rotation in the $(\sigma_1,\sigma_2)$-plane.
The invariance of $\mathrm{d}\Omega_3^2$ under $SU(2)_L\times SU(2)_R$ can be easily checked by using Eq.~(\ref{LRsigma}).

The generators $L_i$ and $R_i$ share the same Casimir operator: $L^2\equiv  L_1^2+L_2^2+L_3^2=R_1^2+R_2^2+R_3^2$, and the set of commutative operators is given by ($L^2$, $L_z$, $R_z$).
The Wigner D-matrix $D^j_{mk}(\theta,\phi,\chi)$ is defined to be the eigenfunction of these operators:
\begin{equation}
L^2 D^j_{mk} = j(j+1) D^j_{mk}\ ,\quad
L_z D^j_{mk} = m D^j_{mk}\ ,\quad
R_z D^j_{mk} = k D^j_{mk}\ ,
\label{Deigen}
\end{equation}
where the ranges of the quantum numbers $(j,m,k)$ are
\begin{equation}
\begin{split}
 j&= 0, \, 1/2, \, 1,\, 3/2, \ldots \ , \\
m&= -j, \, -j+1,\ldots, \, j \ , \\
k&= -j, \, -j+1, \ldots, \, j \ .
\end{split}
\label{jmkrange}
\end{equation}
The Wigner D-matrices are orthogonal under the inner product \eqref{innerprod},
\begin{equation}
 (D^{j'}_{m'k'},D^j_{mk})=\frac{2\pi^2}{2j+1}\delta_{mm'}\delta_{kk'}\delta_{jj'}\ .
\label{Dorth}
\end{equation}

We can also define the ladder operators $L_{\pm} = L_x \pm i L_y$ and $R_{\pm} = R_y \pm i R_x$, which shift the ``orbital angular momenta'' of $D^j_{mk}$ as
\begin{equation}
 \begin{split}
L_+ D^j_{mk} &= \varepsilon_{m+1} D^j_{(m+1)k}\ , \quad
L_- D^j_{mk} = \varepsilon_{m} D^j_{(m-1)k}\ , \\
R_+ D^j_{mk} &= \epsilon_{k+1} D^j_{m(k+1)}\ , \quad \;
R_- D^j_{mk} = \epsilon_{k} D^j_{m(k-1)}\ ,
 \end{split}
\label{Dladder}
\end{equation}
where $\varepsilon_m=\sqrt{(j+m)(j-m+1)}$ and $\epsilon_k=\sqrt{(j+k)(j-k+1)}$.

The Wigner D-matrices satisfy a convenient formula for summation,
\begin{equation}
 \sum_{m=-j}^j (D^j_{mk'})^\ast D^j_{mk}=\delta_{k'k}\ .
\label{DmDm}
\end{equation}
This can be proved easily using the ladder operators. Using Eqs.~(\ref{Deigen}) and (\ref{Dladder}), we find $L_i (\sum_{m=-j}^j (D^j_{mk'})^\ast D^j_{mk}) = 0$.
Hence, the left hand side of \eqref{DmDm} is a constant.
Integrating this equation over the $S^3$ as Eq.~(\ref{innerprod}) and using Eq.~(\ref{Dorth}), we find that the constant is $\delta_{kk'}$.

For later convenience, we introduce a $(2j+1)$-component vector $\vec{D}_k$ by
\begin{equation}
 \vec{D}_k=
\begin{pmatrix}
D^j_{m=j, k} \\
D^j_{m=j-1, k} \\
\vdots\\
D^j_{m=-j, k}
\end{pmatrix}
\ .
\end{equation}
Although $\vec{D}_k$ also depends on the index $j$, we suppress it for notational simplicity because we will generally keep $j$ fixed once the content of the scalar field is specified.
In this notation, Eq.~(\ref{DmDm}) is simply written as
\begin{equation}
\vec{D}_{k'}^\ast \cdot \vec{D}_k = \delta_{k'k}\ .
\label{orthD}
\end{equation}

\section{Cohomogeneity-1 spacetime with rotating scalar field}
\label{sec:scalar_coh1}

We now describe the ansatz for resonating cohomogeneity-1 spacetimes. We will show that the energy-momentum tensor of the matter field we introduce is consistent with the symmetries of the metric and therefore that the equations of motion reduce to a consistent set of ordinary differential equations.

We consider the following five-dimensional Einstein-scalar system with a negative cosmological constant:
\begin{equation}
 S=\frac{1}{16\pi G_5}\int \mathrm{d}^5x\sqrt{-g}\left(R+\frac{12}{L^2}-\partial^\mu \vec{\Pi}^\ast \cdot \partial_\mu \vec{\Pi}\right)\ ,
\label{Estnscalar}
\end{equation}
where $\vec{\Pi}$ denotes a $(2j+1)$-component complex scalar multiplet, $G_5$ is the five-dimensional Newton constant, and $L$ is the AdS radius.
Hereafter, we set $L=1$.

For the metric, we take the cohomogeneity-1 ansatz \cite{Ishii:2018oms}
\begin{multline}
 \mathrm{d}s^2=-(1+r^2)f(r)\mathrm{d}\tau^2 + \frac{\mathrm{d}r^2}{(1+r^2)g(r)}\\
+\frac{r^2}{4}
\{
\alpha(r)\sigma_1^2+\frac{1}{\alpha(r)}\sigma_2^2+\beta(r)(\sigma_3+2h(r)\mathrm{d}\tau)^2
\}\;,
\label{SU2metric}
\end{multline}
where $\sigma_i$ were defined in \eqref{inv1form}.
For the scalar multiplet, we take
\begin{equation}
\vec{\Pi}(\tau,r,\theta,\phi,\chi)= \sum_{k\in K} \Phi_k(r) \vec{D}_k(\theta,\phi,\chi)\ ,
\label{PiSU2}
\end{equation}
where $\Phi_k(r)$ are real scalar fields, and $K$ is defined by
\begin{equation}
\begin{split}
 &K=\{j,j-2,j-4,\cdots,-j\}\quad (j:\textrm{integer}), \\
 &K=\{j,j-2,j-4,\cdots,-j+1\}\quad (j:\textrm{half integer})\ .
\end{split}
\label{Kdef}
\end{equation}

We first comment on the metric ansatz \eqref{SU2metric} before later addressing the scalar.  This metric ansatz preserves $SU(2)_L$ but breaks $SU(2)_R$.  If $\alpha(r)=1$, then a $U(1)_R \subset SU(2)_R$ symmetry generated by $R_3$ is restored.  To see this, we can use the fact that
\begin{equation}
\sigma_1^2+\sigma_2^2=\mathrm{d}\theta^2+\sin^2\theta \mathrm{d}\phi^2=\mathrm{d}\Omega_2^2\;,
\end{equation}
which is independent of $\chi$.

In the metric \eqref{SU2metric}, we also assume invariance under two discrete transformations $P_1$ and $P_2$ defined by
\begin{equation}
 P_1(\tau,\chi,\phi)=(-\tau,-\chi,-\phi)\ ,\quad
 P_2(\tau,\chi,\phi)=(\tau,\chi+\pi,\phi)\ .
\label{P1P20}
\end{equation}
The 1-forms $(\mathrm{d}\tau,\sigma_1,\sigma_2,\sigma_3)$ are transformed by $P_1$ and $P_2$ as
\begin{equation}
\begin{split}
&P_1(\mathrm{d}\tau,\sigma_1,\sigma_2,\sigma_3)=(-\mathrm{d}\tau,-\sigma_1,\sigma_2,-\sigma_3), \\
&P_2(\mathrm{d}\tau,\sigma_1,\sigma_2,\sigma_3)=(\mathrm{d}\tau,-\sigma_1,-\sigma_2,\sigma_3)\ .
\end{split}
\label{P1P2}
\end{equation}
Because of the invariance under $P_1$ and $P_2$, cross terms such as $\sigma_1\sigma_2$ do not appear in Eq.\eqref{SU2metric}.

By examining boundary conditions, it turns out that the metric ansatz \eqref{SU2metric} is taken to be in a frame where asymptotic infinity is rotating.  We will search for black holes with a Killing horizon generated by $\partial_\tau$. This condition in turn enforces that $f(r_h)=g(r_h)=0$. For black hole solutions with $\alpha(r)\neq 1$ we must also satisfy $h(r_h)= 0$ (see appendix~\ref{app:bc}).\footnote{If $\alpha(r) \neq 1$, the $U(1)_R$ isometry $\chi \to \chi + \mathrm{const.}$ is broken, and therefore there is no continuous shift of $h(r)$ that does not change the metric.}
Meanwhile, $h(r)$ approaches a constant value $\Omega$ at infinity $r\to\infty$, and the asymptotic form of the metric becomes
\begin{equation}
ds^2\simeq -r^2\mathrm{d}\tau^2 +\frac{\mathrm{d}r^2}{r^2}+\frac{r^2}{4}\{\sigma_1^2+\sigma_2^2+(\sigma_3+2\Omega\mathrm{d}\tau)^2\}\;,
\end{equation}
from which we see that the boundary metric is $R^{(\tau)}\times S^3$, but with rotation in the $\sigma_3$ directions.
This is the rotating frame at infinity.

The meaning of $\Omega$ becomes clear by moving to the non-rotating frame at infinity, which is the natural frame for interpreting conserved charges and other quantities of the black hole.
We can switch to the non-rotating frame by applying the following coordinate transformation:
\begin{equation}
 \mathrm{d}t=\mathrm{d}\tau\ ,\quad \mathrm{d}\psi=\mathrm{d}\chi+2\Omega \mathrm{d}\tau\ .
\label{tpsidef}
\end{equation}
In the new frame, the horizon generator is written as\footnote{We use a canonically normalized angular coordinate $\psi/2\in [0,2\pi)$.}
\begin{equation}
 \frac{\partial}{\partial \tau}=
\frac{\partial}{\partial t} + \Omega \frac{\partial}{\partial (\psi/2)}\ .
\end{equation}
Therefore, $\Omega$ corresponds to the angular velocity of the horizon.
When $\Omega>1$, the norm of $\partial_\tau$, $g_{\tau\tau}$, becomes positive at infinity.
This implies that there is no global time-like Killing vector in the domain of outer communications, and therefore the spacetime is non-stationary for $\Omega>1$.
If $\alpha\neq1$, the components of the new metric transformed from Eq.~\eqref{SU2metric} become explicitly time dependent. Thus our metric can be said to describe time periodic solutions, even though the metric ansatz is cohomogeneity-1.\footnote{See also Ref.~\cite{Garbiso:2020dys} for a five-dimensional cohomogeneity-1 geometry with periodic time dependence in asymptotically Poincar\'{e} AdS space with a $S^1$ direction.}  In our metric ansatz, we can therefore distinguish between time-periodic solutions (with $\alpha\neq1$) and solutions that are stationary (with $\alpha=1$).

Now we comment on the scalar field \eqref{PiSU2}. This ansatz is precisely the ``double stepping'' ansatz ($k$ decreases by 2 in the sum) introduced in the perturbative analysis of black resonators and geons~\cite{Ishii:2020muv}.  Indeed, the modes in $K$ decouple from those of its complement $K^c=\{j,j-1,\cdots,-j\}\setminus K$.  This fact ultimately stems from the discrete isometry $P_2$ of the metric \eqref{SU2metric}.  Under $P_2$, the Klein-Gordon equation for $\vec{\Pi}$ can be decomposed into even and odd parts.  More specifically, because $\vec{D}_k\propto e^{-ik\chi}$, the Wigner D-matrices with $k\in K$ and $k\in K^c$ acquire different phase factors of $\pm 1$.
The Klein-Gordon equations for $k\in K$ and $k\in K^c$ are therefore decoupled.
In this paper, we will consider only $k\in K$.

Note also that we have defined our ansatz with fixed $j$ and $2j+1$ multiplet.  However, a solution with a particular $j$ is also a solution to the theory with larger multiplets than $2j+1$, simply by setting the extra components of the multiplet to zero.  Solutions with different $j$ can therefore be consistently compared with one another.

Finally, we show that the scalar field ansatz \eqref{PiSU2} is consistent with the metric \eqref{SU2metric}, i.e.~the Einstein and Klein-Gordon equations reduce to a consistent set of ODEs.
The Einstein equations from Eq.~(\ref{Estnscalar}) are given by $G_{\mu\nu}-6g_{\mu\nu}=T_{\mu\nu}$, where
the energy-momentum tensor is
\begin{equation}
 T_{\mu\nu}=\mathcal{T}_{(\mu\nu)}-\frac{1}{2}g_{\mu\nu} \mathcal{T}\ ,\quad
\mathcal{T}_{\mu\nu}=\partial_{\mu} \vec{\Pi}^\ast\cdot \partial_{\nu} \vec{\Pi} \ ,\quad
\mathcal{T}=g^{\mu\nu}\mathcal{T}_{\mu\nu}\ .
\label{Tdef}
\end{equation}

To derive the explicit expression of the energy-momentum tensor for Eq.~(\ref{PiSU2}), it is convenient to
introduce 1-forms $\sigma_\pm$ defined by
\begin{equation}
\sigma_{\pm}=\frac{1}{2}(\sigma_1\mp i \sigma_2)=\frac{1}{2}e^{\mp i\chi}(\mp i \mathrm{d}\theta+\sin\theta \mathrm{d}\phi)
\end{equation}
and use the basis $e^a=\{\mathrm{d}\tau,\mathrm{d}r,\sigma_+,\sigma_-,\sigma_3\}$ $(a=\tau,r,+,-,3)$.
Their dual vectors are given by $e_a=\{\partial_\tau,\partial_r,e_+,e_-,\partial_\chi\}$ where
\begin{equation}
e_\pm= \bar{\xi}_1\pm i\bar{\xi}_2
=\pm R_{\mp}
=e^{\pm i \chi}(\pm i \partial_\theta + \frac{1}{\sin\theta}\partial_\phi - \cot\theta \partial_\chi)\ .
\end{equation}
One can check that $e^a_\mu e_b^\mu=\delta_b^a$.

In this basis, the derivatives of the scalar field can be evaluated by using Eqs.~(\ref{Deigen}) and (\ref{Dladder}) as
\begin{equation}
\begin{split}
&\partial_\tau \vec{\Pi} =0\ ,\quad
\partial_r \vec{\Pi} =\sum_{k\in K}\Phi'_k \vec{D}_k\ ,\quad
\partial_+ \vec{\Pi} =\sum_{k\in K}\epsilon_k \Phi_k \vec{D}_{k-1}\ ,\\
&\partial_- \vec{\Pi} =-\sum_{k\in K}\epsilon_{k+1} \Phi_k \vec{D}_{k+1}\ ,\quad
\partial_3 \vec{\Pi} =-i\sum_{k\in K}k \Phi_k \vec{D}_k\ ,
\end{split}
\label{Pider}
\end{equation}
where $\partial_a \equiv e_a^\mu\partial_\mu$.

Some components of $\mathcal{T}_{ab}$
vanish because of the double stepping coupling~(\ref{Kdef}). For example, we find
\begin{equation}
 \mathcal{T}_{r+}
= \sum_{k,k'\in K}\epsilon_k \Phi'_{k'} \Phi_k  \vec{D}_{k'}^\ast \cdot \vec{D}_{k-1}
=\sum_{k,k'\in K}\epsilon_k \Phi'_{k'} \Phi_k  \delta_{k',k-1}=0\ ,
\end{equation}
where in the second equality we used Eq.~(\ref{orthD}), and the last one follows from the fact that $k'$ and $k-1$ cannot be equal because of the double stepping of $k$~(\ref{Kdef}).

To evaluate the non-vanishing components of $\mathcal{T}_{ab}$, we also use the orthogonality of the Wigner D-matrices (\ref{orthD}).
The upshot is that the energy momentum tensor is given by
\begin{multline}
 \mathcal{T}_{(ab)} e^a e^b
= \sum_{k\in K}\bigg[
\Phi'_{k}{}^2 \mathrm{d}r^2
-\epsilon_{k-1}\epsilon_k \Phi_{k-2}\Phi_k (\sigma_+^2+\sigma_-^2)\\
+(\epsilon_{k}^2+\epsilon_{k+1}^2) \Phi_{k}^2 \sigma_+ \sigma_-
+k^2 \Phi_{k}^2 \sigma_3^2
\bigg]\ .
\label{Tab}
\end{multline}
This result, invariant under $SU(2)_L$, is consistent with the spacetime~(\ref{SU2metric}), and the Einstein and Klein-Gordon equations reduce to a consistent set of coupled ODEs.
The explicit form of the equations of motion is summarized in appendix~\ref{tech}, where technical details in solving the equations are also explained.

In our ansatz of the scalar field~(\ref{PiSU2}), 
the conserved current of the complex scalar field $J_\mu$ is given by 
\begin{equation}
 J_\mu \mathrm{d}x^\mu \equiv \textrm{Im}[\,\vec{\Pi}\cdot \partial_\mu \vec{\Pi}^\ast\,] \mathrm{d}x^\mu
=\sum_{k\in K}k^2\Phi_k^2 \sigma_3\ .
\end{equation}
We have $J_\psi \neq 0$ if and only if $\Phi_k\neq 0$ for $k\neq 0$.
This indicates that there is a rotating flow of the scalar field and 
it carries angular momentum when the scalar field has non-trivial $\psi$-dependence.

\section{Geons and superradiant instability of Myers-Perry-AdS}
\label{sec:geonMPAdS}

\begin{figure}
\begin{center}
\includegraphics[scale=0.5]{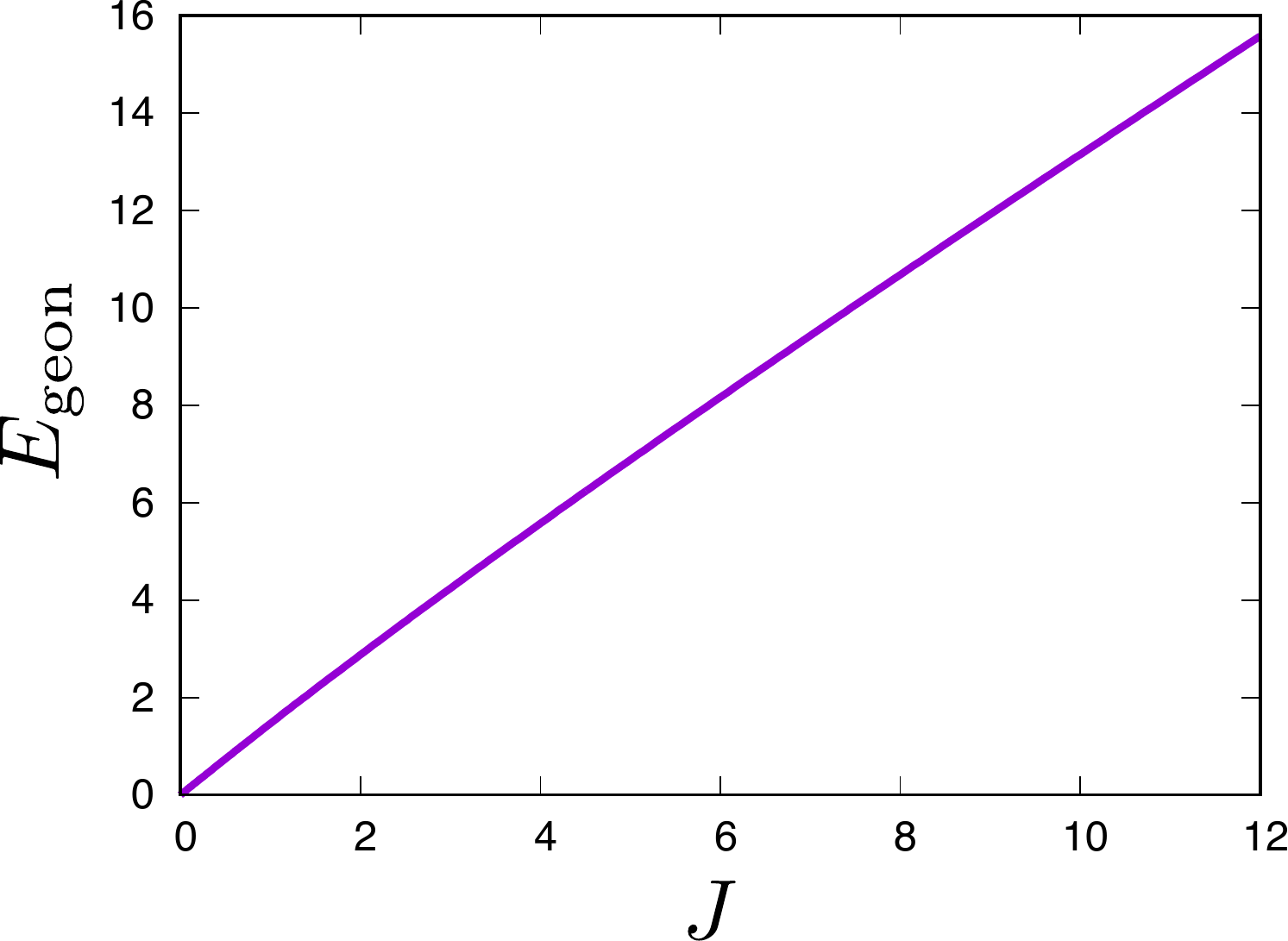}
\end{center}
\caption{Energy of geons $E_\textrm{geon}$ as a function of the angular momentum $J$.
}
\label{Egeon}
\end{figure}

If $\alpha(r)=1$ in Eq.~(\ref{SU2metric}), there is an exact solution describing a rotating black hole with both angular momenta set equal.  It is part of the Myers-Perry-AdS family of solutions, which we will abbreviate MPAdS.  In our ansatz, the metric functions are
\begin{equation}
\begin{split}
&g(r)=1-\frac{2\mu (1-a^2)}{r^2(1+r^2)} +\frac{2a^2\mu}{r^4(1+r^2)}\ ,\quad
\beta(r)=1+\frac{2 a^2\mu}{r^4}\ ,\\
&h(r)=\Omega-\frac{2\mu a}{r^4+2 a^2\mu}\ ,\quad
f(r)=\frac{g(r)}{\beta(r)}\ ,\quad
\alpha(r)=1\ .
\end{split}
\label{MPAdSFunctions}
\end{equation}
The event horizon $r=r_h$ is located at the largest root of $g(r_h)=0$, and the isometry group of this solution is $R^{(\tau)} \times SU(2)_L \times U(1)_R$.

The solution is parametrised by $\mu$ and $a$, with $\mu=0$ corresponding to pure AdS.
As written, $\Omega$ is merely a gauge parameter that allows us to move between rotating and non-rotating frames at infinity.
For consistency of notation and convenience, we set
\begin{equation}
 \Omega=\frac{2\mu a}{r_h^4+2 a^2\mu}
\end{equation}
so that $\Omega$ is the angular velocity of the horizon. For this choice, we have $h(r_h)=0$ and, thus, $\partial_\tau$ becomes the horizon generator.  MPAdS solutions are bounded by extremality, which occurs at
\begin{equation}
 \Omega_\mathrm{extr}=\frac{\sqrt{1+2r_h^2}}{\sqrt 2 r_h}\,.
\end{equation}

Before discussing perturbations of MPAdS, let us briefly describe a family of solutions called geons.  Geons are horizonless, nonlinear extensions of gravitational normal modes of pure AdS.  Within our ansatz, these normal modes are given by a perturbation of the form $\alpha=1+\delta\alpha(r)$ about AdS, with $\Omega$ appearing as an eigenvalue.  Geons therefore carry angular momentum, with $\Omega$ as an angular frequency.  Fig.~\ref{Egeon} is the energy of geons $E_\textrm{geon}$ as a function of the angular momentum $J$.  Later in this paper, we will primarily use the difference of the energy $E$ from $E_\textrm{geon}$ in figures for better visibility. Near vacuum AdS, the geons have frequency parameter $\Omega\simeq 3/2$, and hence the energy scales as $E_\textrm{geon}\simeq (3/2)J$.

\begin{figure}
  \centering
\subfigure
 {\includegraphics[scale=0.5]{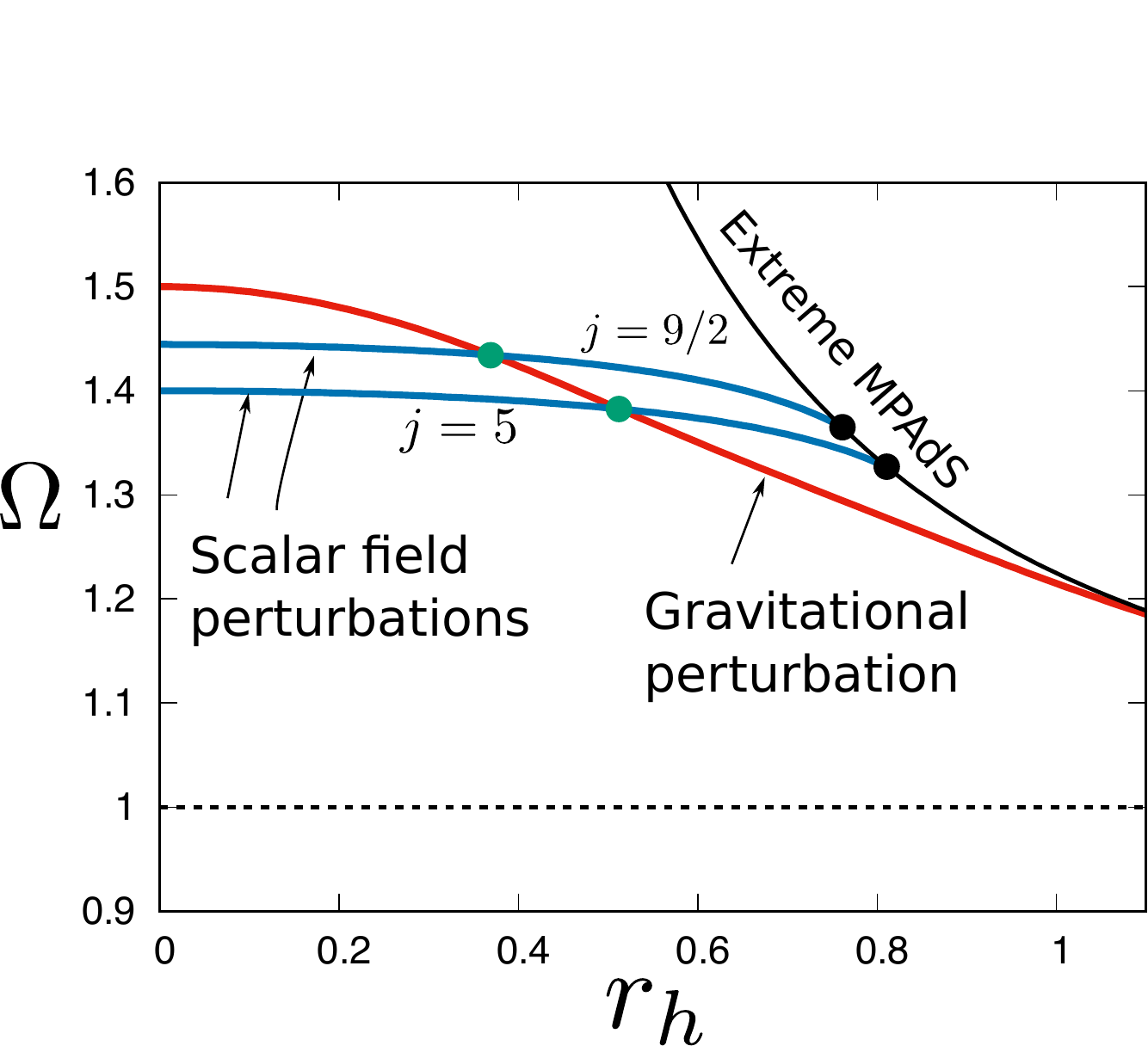}\label{onset_Om}
  }
  \subfigure
 {\includegraphics[scale=0.5]{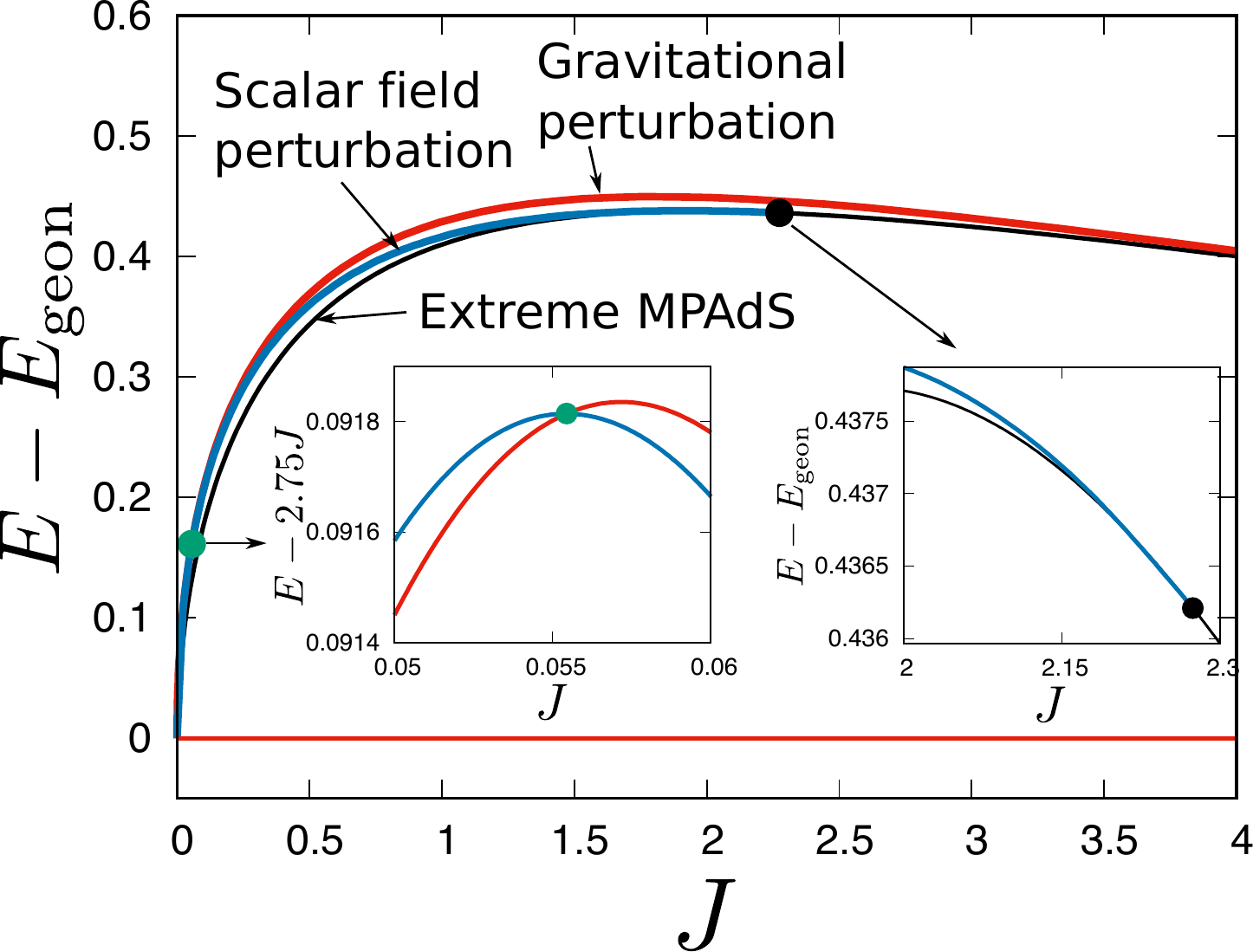}\label{onset_EJ}
  }
 \caption{
(Left) Onset of the superradiant instability of MPAdS against gravitational perturbation (red) and scalar field perturbations with $j=k=9/2,\,5$ (blue) in the $(\Omega,r_h)$-plane. Above the red and blue curves,
MPAdS is unstable to gravitational and scalar perturbations, respectively.
The extreme limit of MPAdS is shown by the black curve.
(Right) The same data as the left panel are shown in the $(E-E_\textrm{geon},J)$-plane, though for the scalar field perturbations only the onset of the mode with $j=k=9/2$ is shown for visibility.
}
\label{onset}
\end{figure}

Now let us return to perturbations of MPAdS black holes.  When the MPAdS black hole has sufficiently high angular frequency, it is unstable to superradiance against both gravitational and scalar field perturbations.
Among the unstable superradiant modes is one that breaks only the $U(1)_R$ isometry of the metric.
Its onset mode of the gravitational perturbation can be found in our ansatz by perturbing the metric function as $\alpha(r)=1+\delta\alpha(r)$ about MPAdS and linearising the equation of motion~(\ref{EOMa}) in $\delta\alpha(r)$.
The onset mode of the scalar field perturbation is just given by the probe scalar field satisfying the Klein-Gordon equation~(\ref{EOMa}).
In the MPAdS background, modes with different $k$ decouple in the Klein-Gordon equation.  We will focus on modes with $k=j$, as these are the most dominant.

In Fig.~\ref{onset}, we show results for gravitational and scalar field onsets with $j=k=9/2$ and $j=k=5$. In the left figure, we show the onset of the instabilities in the $(\Omega,r_h)$- and $(E-E_\textrm{geon},J)$-planes.  The scalar and gravitational onset curves intersect at green dots.  For reference, we also show where MPAdS is extremal, and where $\Omega=1$.  Recall that MPAdS is unstable for $\Omega>1$, but not necessarily to the $j=k=9/2$ and $j=k=5$ modes.

In the right figure, we show the same results in the $(E-E_\textrm{geon},J)$-plane, 
but only show the result for the scalar for $j=k=9/2$ for visibility.
The onset of the instability for gravitational and scalar field perturbations are shown by red and blue curves, respectively.
These onset curves intersect at green dots.
The extreme MPAdS is also shown by the black curve.
Onset curves of scalar field perturbations terminate at black dots on the extreme MPAdS.
In the insets in the right panel, we zoom in on the regions around the green and black dots for visibility.
Note that, in the left inset, we take $E-2.75J$ as the vertical axis for visibility.

\section{Black resonators}
\label{sec:BR}

As mentioned in the previous section, MPAdS is unstable to superradiance against gravitational perturbations.
A new family of cohomogeneity-1 solutions with $\alpha(r)\neq 1$ branches off from the onset of the instability \cite{Ishii:2018oms}.  These onsets were given by the red curves labelled ``gravitational perturbations'' in Fig.~\ref{onset}. Because $\alpha\neq1$ for this new family, these black holes are time-periodic as seen in the non-rotating frame at infinity and are known as black resonators.  These black resonators have $R \times SU(2)_L$ isometries.  Like MPAdS black holes, these black resonators are a two-parameter family.  The horizonless limit of black resonators are geons \cite{Dias:2011ss,Horowitz:2014hja,Martinon:2017uyo,Fodor:2017spc}, which we have already described in the previous section and in Fig.~\ref{Egeon}.  Like black resonators, geons also have $R \times SU(2)_L$ isometries and are time-periodic in the non-rotating frame at infinity.

\begin{figure}
\begin{center}
\includegraphics[scale=0.6]{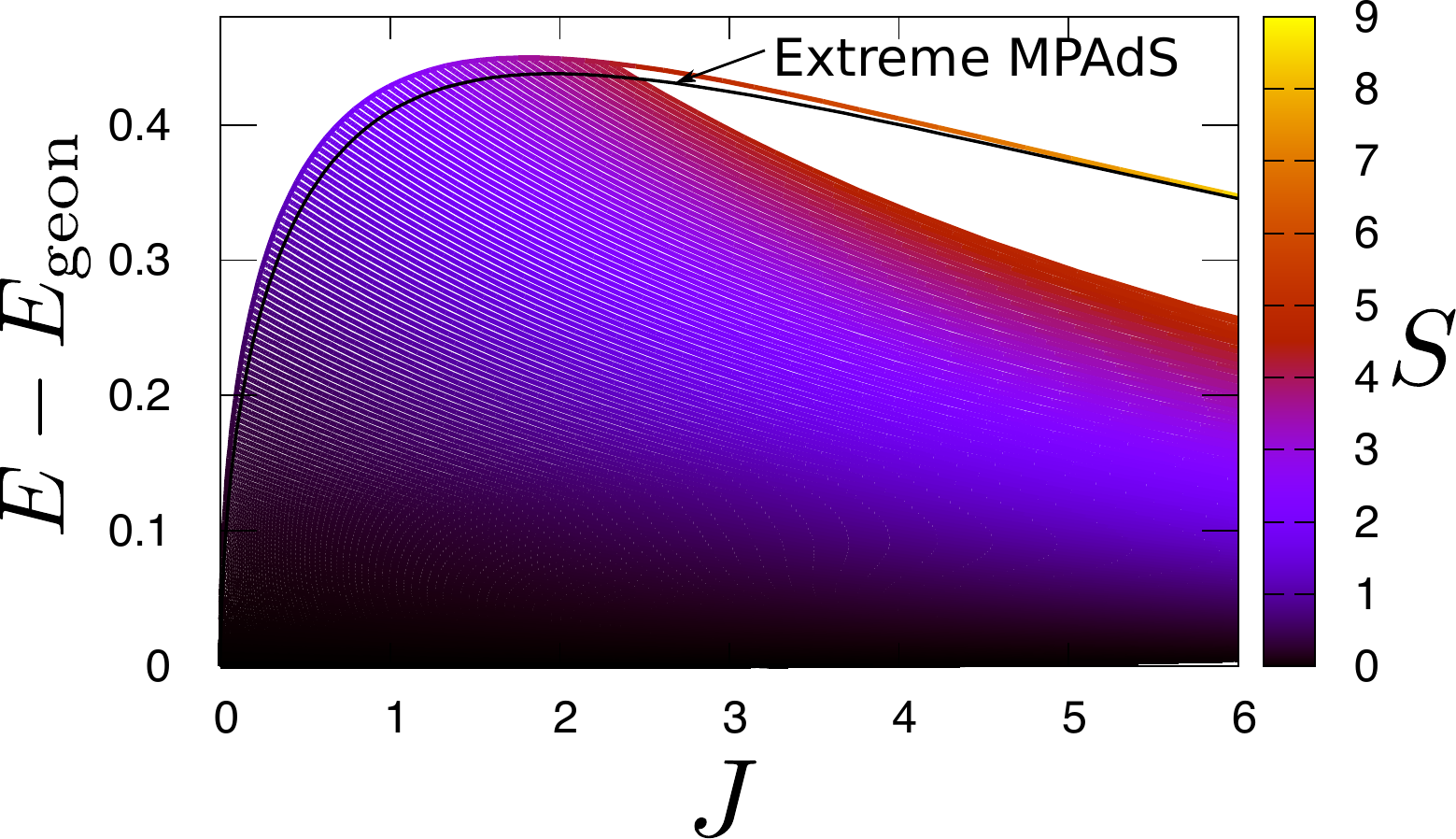}
\end{center}
\caption{
Entropy of black resonators as a function of $(E-E_\textrm{geon},J)$.
The upper solid curve indicates the onset of the gravitational superradiant instability of the MPAdS, where the black resonators branch off. The extreme MPAdS is shown by a black curve. In the upper-right region below the onset curve, we simply do not have numerical data.
}
\label{Sbr}
\end{figure}

In Fig.~\ref{Sbr}, the entropy of black resonators $S$ is shown by the colour map as a function of $(E-E_\textrm{geon},J)$.
The extreme MPAdS is shown by a black curve. Only MPAdS exists in the upper side of the black curve.
The black resonators branch off from the onset of the superradiant instability shown by the solid curve on the upper edge of the plotted region. In the upper-right white region below the onset curve, we do not have numerical data for black resonators.
For a large $J$, the MPAdS at the onset is very close to the extremality, and numerical construction of the black resonator becomes difficult.
In the limit of geons, $E-E_\textrm{geon} \to 0$, the entropy also approaches zero.
In Ref.~\cite{Ishii:2018oms}, it has been found that the angular velocity of the black resonator and geon always satisfied $\Omega>1$.
Therefore, the black resonator and geon are superradiant and non-stationary.

\section{Hairy Myers-Perry-AdS and boson stars}
\label{sec:HMPAdS}

Thus far, we have discussed MPAdS, black resonators, and geons, which are all solutions that satisfy $\vec{\Pi}=0$.  Now we turn to solutions with $\vec{\Pi}\neq0$, beginning with those with $\alpha=1$.

A special case of our ansatz~(\ref{PiSU2}) is given by the scalar field with ``single-$k$'':
\begin{equation}
 \vec{\Pi}(\tau,r,\theta,\phi,\chi) = \Phi_k(r)\vec{D}_{k}(\theta,\phi,\chi) \quad (\textrm{no summation})\ .
\label{Pisinglek}
\end{equation}
In this case, the matter stress tensor Eq.~(\ref{Tab}) reduces to
\begin{equation}
 \mathcal{T}_{(ab)} e^a e^b
= \Phi'_{k}{}^2 \mathrm{d}r^2 +\frac{1}{4}(\epsilon_{k}^2+\epsilon_{k+1}^2) \Phi_{k}^2 (\sigma_1^2+\sigma_2^2)
+k^2 \Phi_{k}^2 \sigma_3^2\ ,
\label{Tab2}
\end{equation}
where we returned to $\sigma_{1,2}$  from $\sigma_\pm$.
In this expression, the coefficients of $\sigma_1$ and $\sigma_2$ coincide, and therefore it has the invariance under $U(1)_R \in SU(2)_R$ generated by the angular momentum operator $R_3$.
The metric (\ref{SU2metric}) therefore also has this isometry, which implies that $\alpha(r)=1$ in the single-$k$ case.

\begin{figure}
  \centering
\subfigure[Energy of boson star]
 {\includegraphics[scale=0.45]{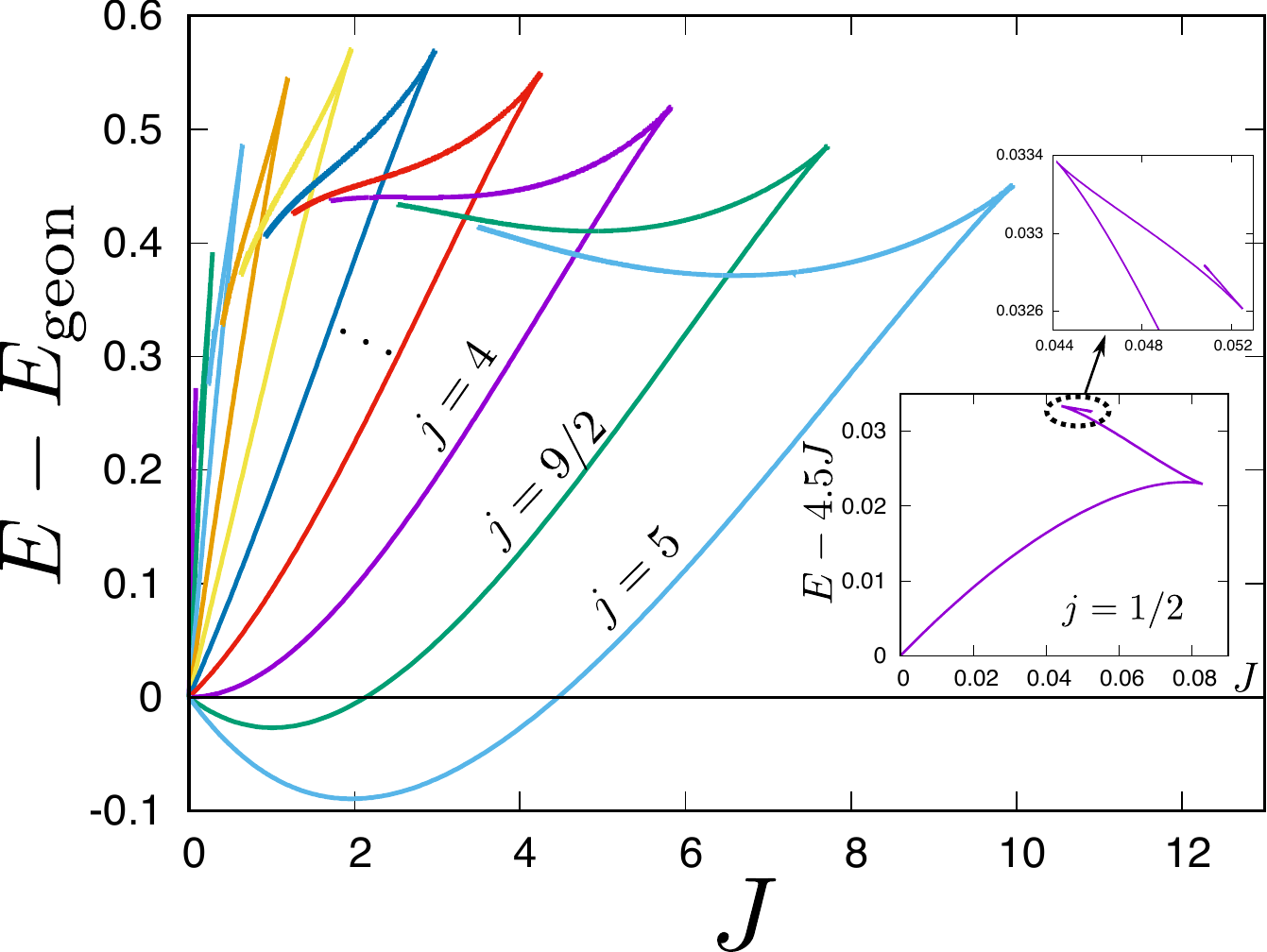}\label{Ebs}
  }
  \subfigure[Entropy of hairy MPAdS]
 {\includegraphics[scale=0.55]{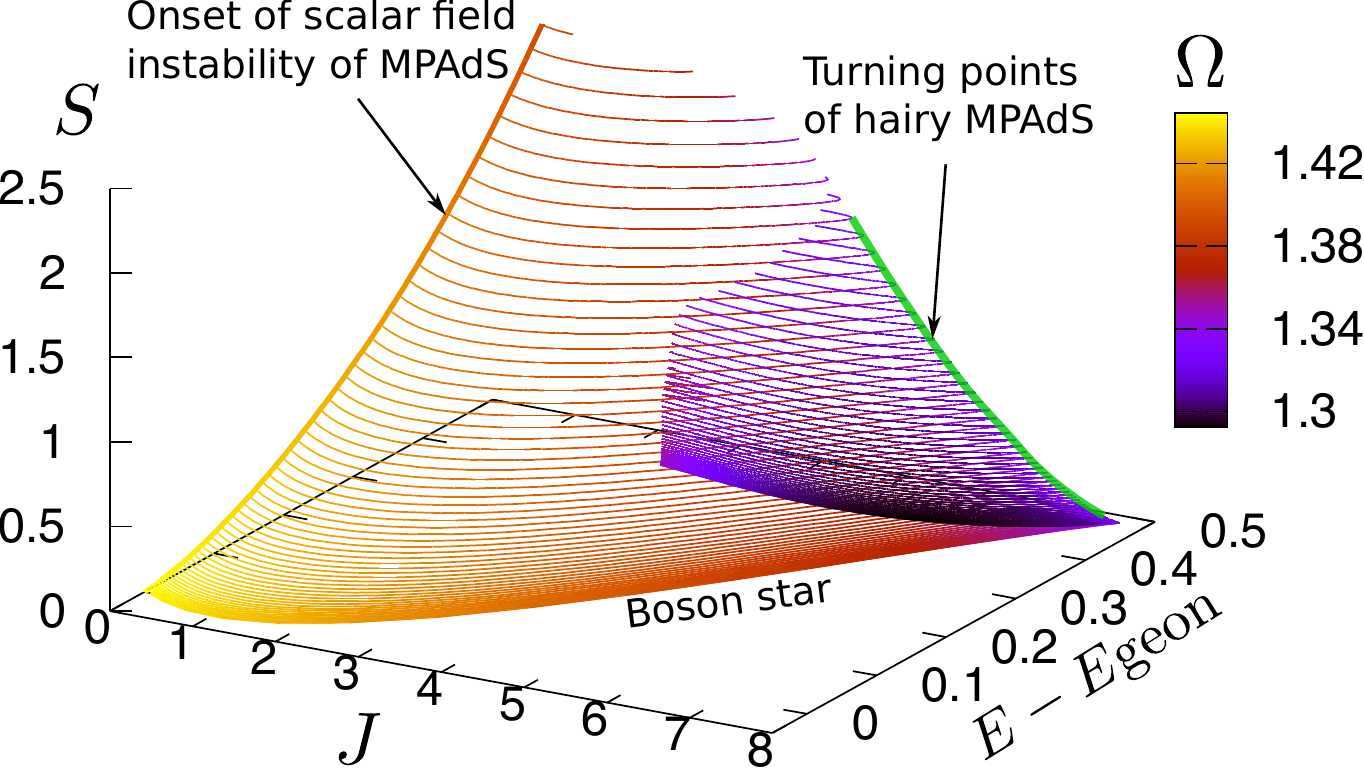}\label{SHMPAdS}
  }
\subfigure[Domain of hairy MPAdS]
 {\includegraphics[scale=0.5]{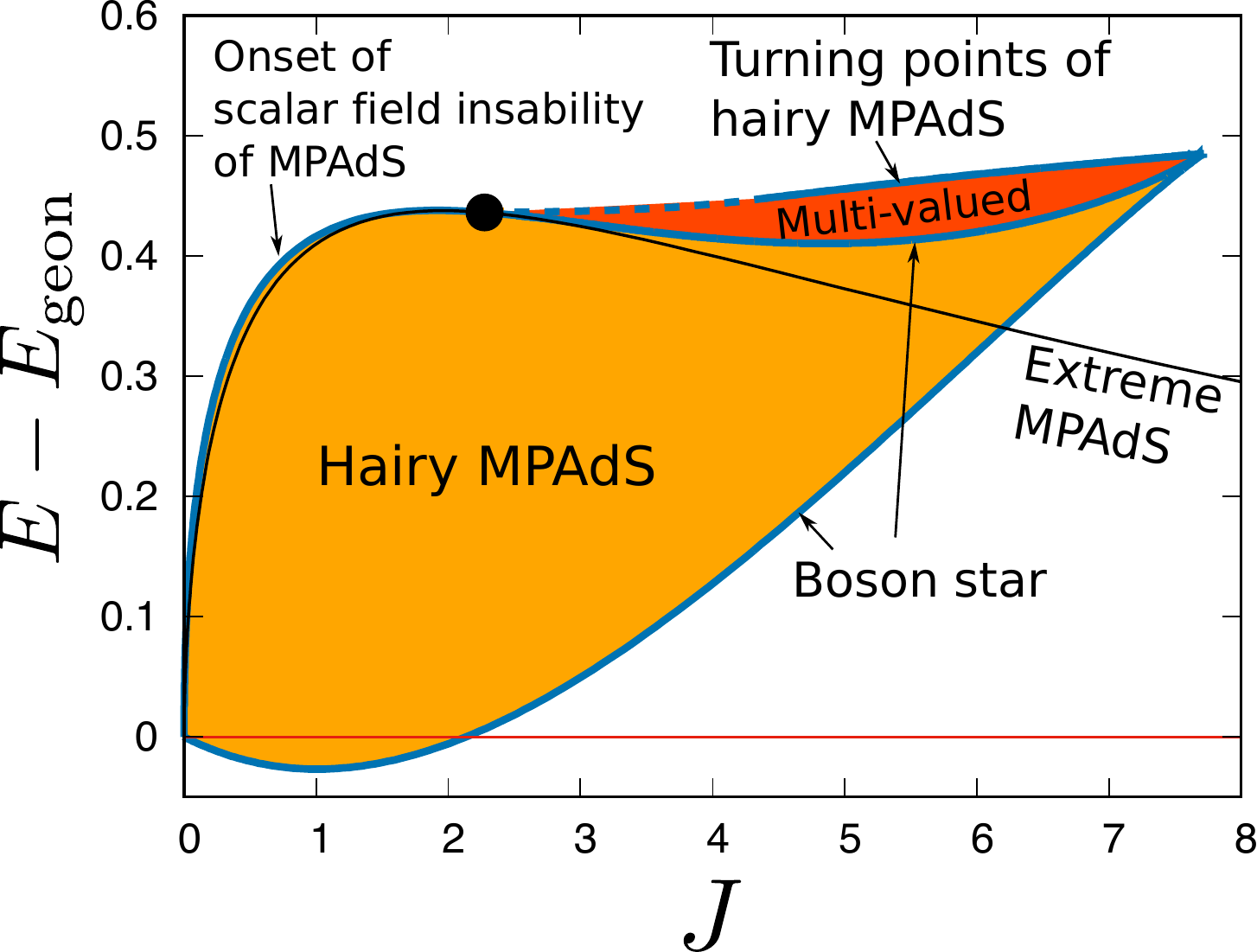}\label{HMPAdSdomain}
  }
 \caption{
(a) Energy of boson stars as a function of the angular momentum $J$.
The existence of multiple turning points are shown in the insets for $j=1/2$.
(b) Entropy of the hairy MPAdS for $k=j=9/2$, parametrized by $J$ and $E-E_\textrm{geon}$.
The solid curve on the left edge indicates the onset of the scalar field superradiant instability of the MPAdS, from which the hairy MPAdS branches off.
(c) Domain of existence of the hairy MPAdS. In orange and red regions, the hairy MPAdS exists. In the red region, the entropy and other physical quantities become multi-valued because of the turning points.
}
\end{figure}

Superradiant instabilities can be induced by scalar fields as well as gravitational fields.  These onset curves were shown earlier for the scalar field with $j=k=9/2$ and $j=k=5$ by the blue curves in Fig.~\ref{onset}. For the scalar field, we refer to the solutions branching off from the onset of the superradiant instability as the \textit{hairy MPAdS black holes}, as they contain scalar hair.  Because black resonators can be interpreted as black holes with gravitational hair, hairy black holes are the scalar field counterparts to black resonators.

A hairy MPAdS black hole becomes a boson star in the horizonless limit.  Boson stars can also be described as nonlinear scalar normal modes of pure AdS.  Boson stars are therefore the scalar field counterparts to geons.  We obtain the perturbative solution of the boson star near pure AdS in appendix~\ref{largej}, where we also discuss the large-$j$ limit for the perturbative solution.
In the following, we focus on the scalar field with $k=j$, which is the most relevant mode for the superradiant instability for a given $j$.
We note that the hairy MPAdS black hole with $j=1/2$ was constructed in Ref.~\cite{Dias:2011at}.\footnote{
The ansatz~(\ref{Pisinglek}) for $j=1/2$ was first considered in Ref.~\cite{Hartmann:2010pm} for constructing rotating boson stars in asymptotically flat spacetime.}
Our treatment (\ref{Pisinglek}) gives a generalization to $j\geq 1$.

In Fig.~\ref{Ebs}, the energy of boson stars are shown as a function of the angular momentum $J$.
The curves correspond to $j=1/2,1,3/2,\cdots,5$ from left to right.
The difference of the energy from that of the gravitational geon (see Fig.~\ref{Egeon}) is used in the vertical axis.

These results near small $E$ and $J$ agree with perturbation theory about pure AdS.  More specifically, the scalar field (\ref{Pisinglek}) with $k=j$ has the lowest normal mode at $\Omega=1+2/j$, from which the boson star branches off.\footnote{The normal mode frequency for the scalar field with the quantum numbers $(j,k)$ is given by $\Omega = (2+j+n)/k$ where $n$ is the radial overtone number. The lowest mode has $n=0$. One can also see that $k=j$ gives the lowest $|\Omega|$.}
For small $J$, the energy of the boson star is given by $E\simeq(1+2/j) J$.
Comparing this with that of the gravitational geon $E_\textrm{geon}\simeq (3/2)J$, one finds that $E-E_\textrm{geon}$ in small $J$ is $\mathcal{O}(J^2)$ for $j=4$ and negative for $j \ge 9/2$.

As $J$ is increased, there are turning points in the energy of boson stars, indicating the change of stability of the boson stars.
It is common for solutions past the turning points to be unstable \cite{poincare1885,Sorkin:1981jc,Sorkin:1982ut,Arcioni:2004ww}.
In the insets of the figure, we zoom in on the curve for $j=1/2$ 
(here $E-4.5J$ is used in the vertical axis for visibility).
The curve is folded multiple times. 

The entropy of the hairy MPAdS for $j=9/2$ is shown in Fig.~\ref{SHMPAdS}.
The angular velocity of the horizon is also shown by the colour map.
For the result of $j=5$, see section \ref{sec:phasej5}.
The hairy MPAdS branches off from the onset of the scalar field superradiant instability denoted by the solid curve on the left edge of the plot region.
Turning points of the energy also exist for hairy MPAdS as shown by the green curve,
and as a result the entropy and other physical quantities of the hairy MPAdS become multi-valued.
We see $\Omega>1$, and this indicates that the hairy MPAdS and boson star are superradiant and non-stationary.

Fig.~\ref{HMPAdSdomain} shows the domain of existence of the hairy MPAdS in the ($E-E_\textrm{geon},J$)-plane. In orange and red regions, the hairy MPAdS exists.
In particular, in the red region, there are multiple solutions for a fixed $(E,J)$, and physical quantities are multi-valued.
Our numerical calculation indicates that the curve of turning points terminates at the black dot: the intersecting point between the extreme MPAdS and the onset of scalar field superradiant instability of MPAdS.
A part of the turning points is shown by the dashed blue curve.
We drew this part by interpolation of our numerical data.

\section{Hairy black resonator and graviboson star}
\label{sec:HBR}

Finally, we consider the most general case in our ansatz (\ref{SU2metric}) and (\ref{PiSU2}) that has both $\alpha(r) \neq 1$ and $\vec{\Pi}(r) \neq 0$.

Ref.~\cite{Ishii:2020muv} has located the onset of the superradiant instability of black resonators for scalar fields.
At the onset of the instability, there is a $\tau$-independent perturbation of the scalar field in the form of Eq.~(\ref{PiSU2}). This is expected to lead to a new family of black resonator solutions with a nontrivial scalar hair.

Similarly, the hairy MPAdS black hole is expected to be unstable against gravitational perturbations with $\alpha(r)\neq 1$ (see appendix~\ref{InstHMPAdS}), and a new family of black resonator solutions is expected to branch from the onset of this instability.

We can therefore have scalar hair on black resonators, and gravitational ``resonator''-type excitations on hairy MPAdS black holes. It turns out both of these excitations are part of the same family of solutions, which we call \textit{hairy black resonators}. Hairy black resonators have their own family of horizonless solutions, which we refer to as \textit{graviboson stars}, as they resemble a combination of a geon and a boson star.

\begin{figure}
  \centering
\subfigure[Domain of hairy black resonator]
 {\includegraphics[scale=0.5]{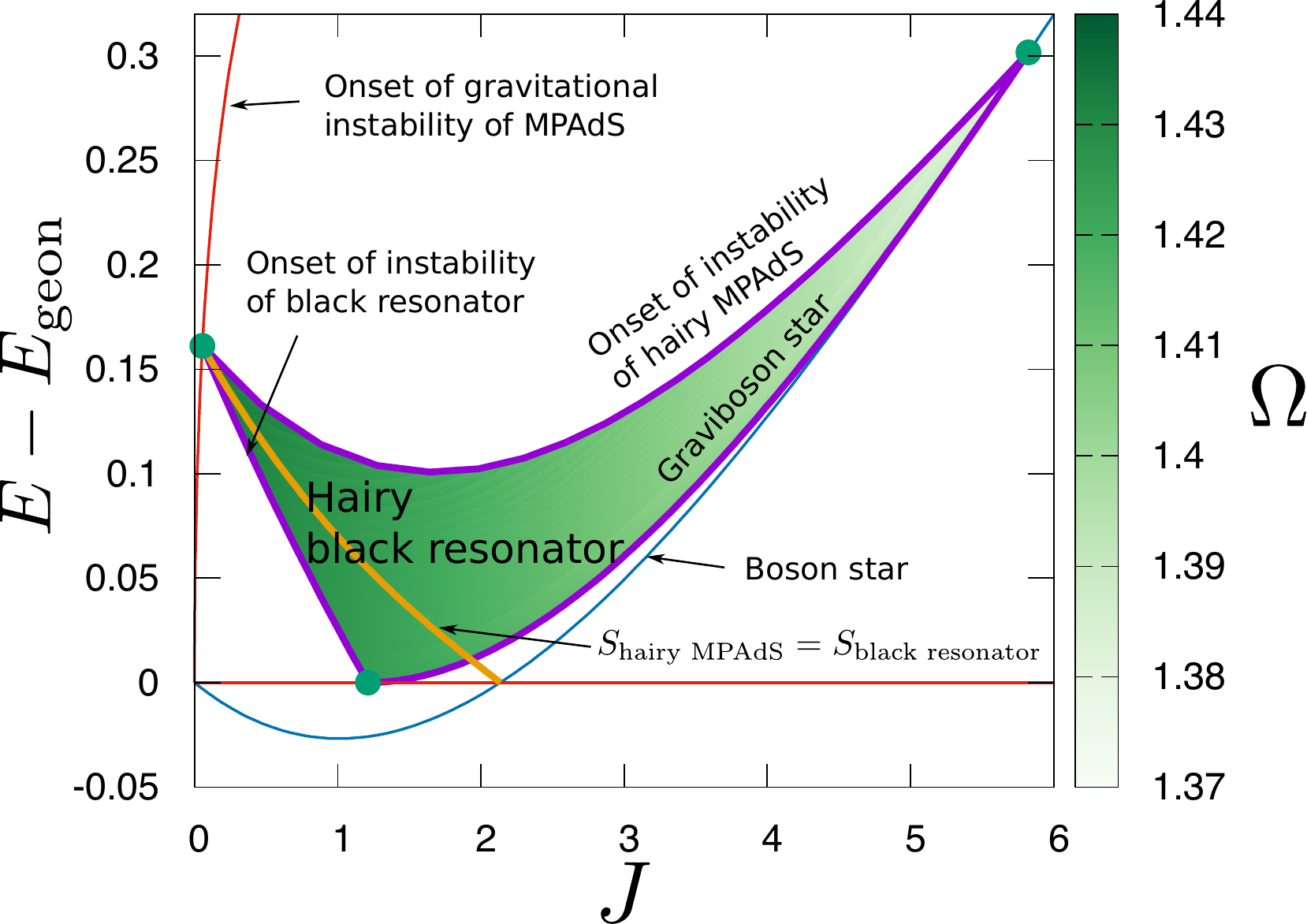}\label{trig}
  }
\subfigure[Entropy]
 {\includegraphics[scale=0.5]{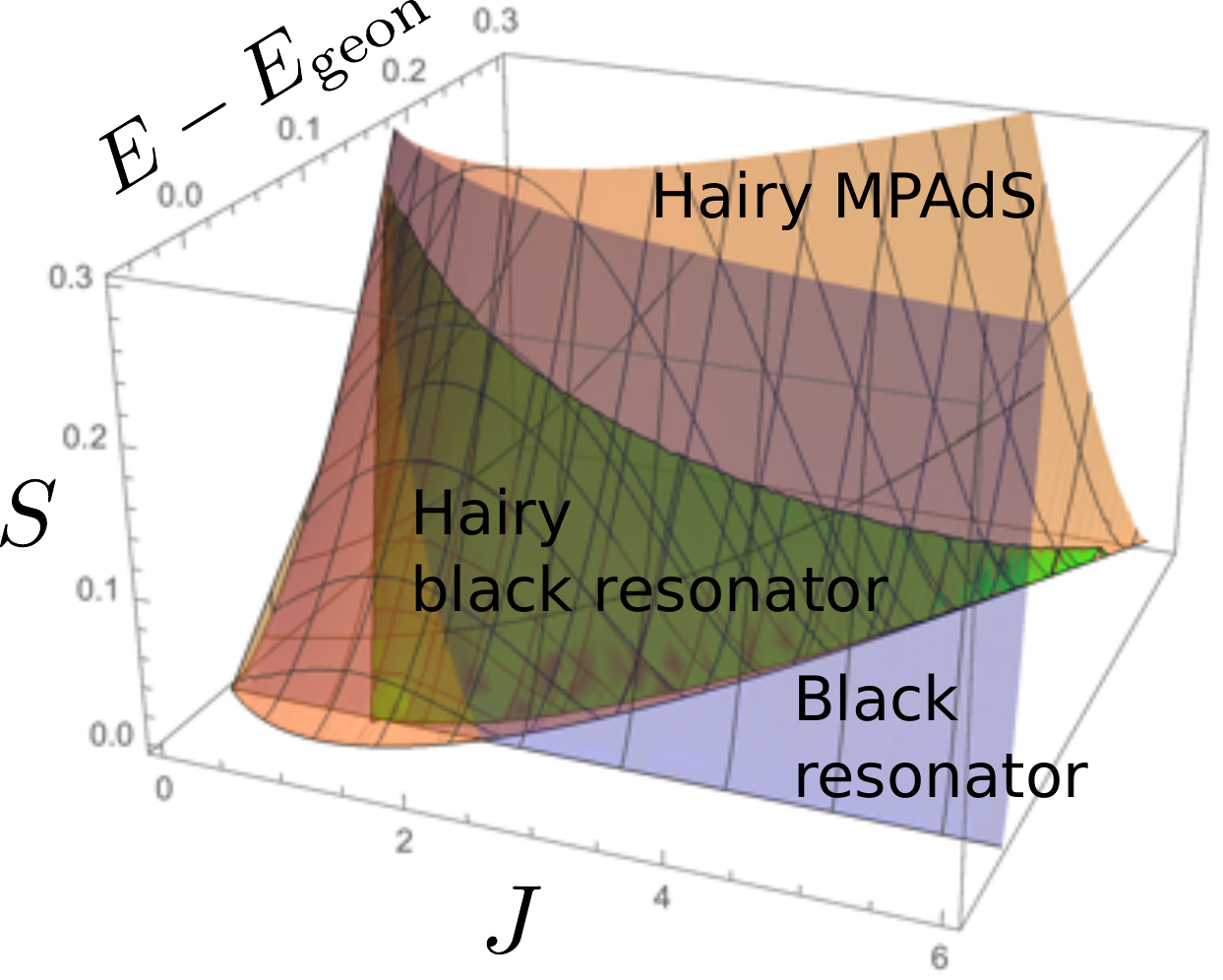}\label{S3sol}
  }
  \subfigure[Cross section at $J=0.8$]
 {\includegraphics[scale=0.6]{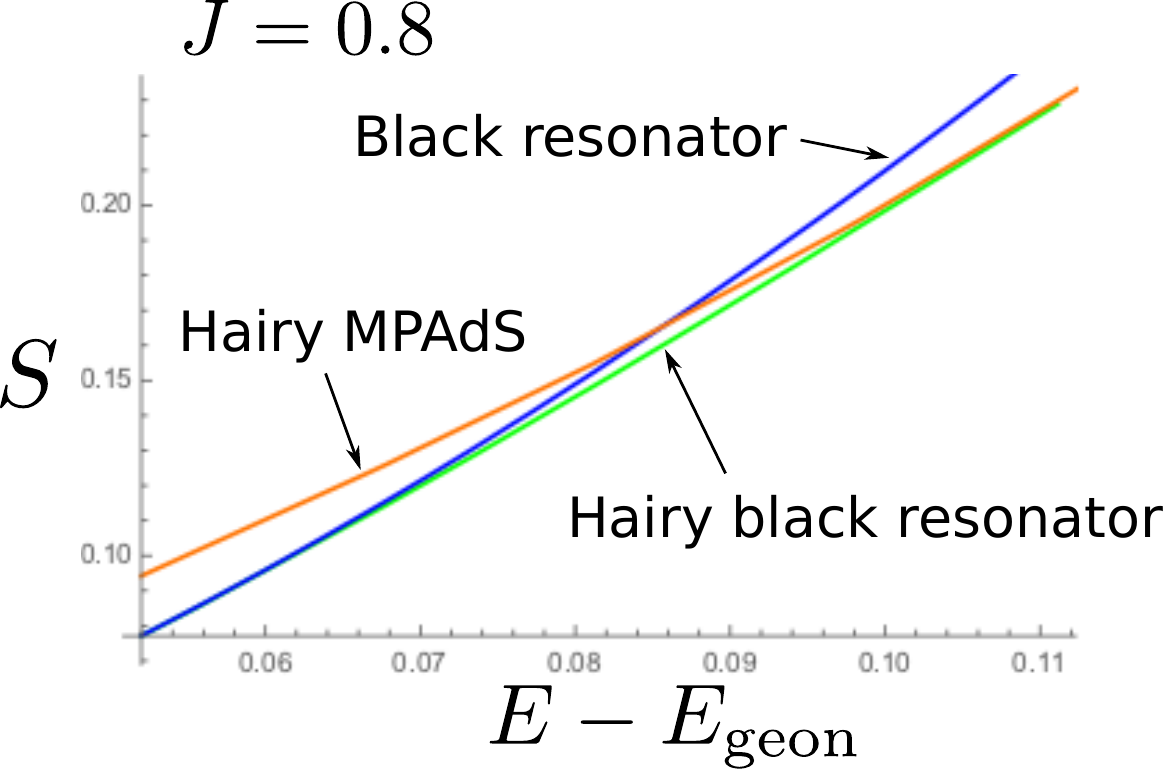}\label{S3sol_Jfix}
  }
 \caption{
(a) Domain of existence of the hairy black resonator. It exists inside the ``triangle'' surrounded by purple curves.
The equal-entropy-curve between the hairy MPAdS and black resonator is plotted by the orange curve.
The angular velocity is shown by the color map. 
(b) Entropies of the black resonator (blue), hairy MPAdS (orange), and hairy black resonator (green) for $j=9/2$.
The entropy of the hairy black resonator is never the largest.
(c) Cross section of (a) at $J=0.8$.
}
\end{figure}

Fig.~\ref{trig} shows the domain of existence of the hairy black resonator.
The hairy black resonator exists in the coloured ``triangular'' domain surrounded by the purple curves.
The colour map corresponds to the angular velocity of the horizon.
The top and bottom-left edges of the triangle correspond to the onset of instability of the hairy MPAdS and black resonator, respectively.
The hairy MPAdS is unstable in the upper side of the top purple curve.
The black resonator is unstable in the lower side of the bottom-left purple curve.
The bottom-right edge is the horizonless limit of the hairy black resonator: graviboson star.
We also show the equal-entropy-curve between the hairy MPAdS and black resonator by the orange curve.
The angular velocity of the hairy black resonator always satisfies $\Omega>1$. Therefore, the hairy black resonator is also superradiant and non-stationary.

We can now compare the entropy of the hairy black resonators to that of black resonators and hairy black holes. Fig.~\ref{S3sol} gives a summary of the entropies of the black resonator (blue), hairy MPAdS (orange), and hairy black resonator (green) for $j=9/2$, which is the smallest $j$ for which black resonators can be unstable. Fig.~\ref{S3sol_Jfix} corresponds to its slice at $J=0.8$.
We find that the entropy of the hairy black resonator is never the largest among the available solutions.  Instead, the most entropic solution is either a hairy MPAdS black hole or a black resonator (or MPAdS, but only in regions where none of the other solutions exist).  
The entropy changes continuously, but the field configurations are discontinuous across this transition.

The hairy black resonator and graviboson star can be interpreted as a simple example of a multi-oscillating solution.
Recall that the Wigner D-matrix depends on $\chi$ as $D_k(\theta,\phi,\chi) \propto e^{-ik\chi}$.
Then, in the the non-rotating frame at infinity~(\ref{tpsidef}),
the scalar field of the hairy black resonator
can be written as
\begin{equation}
 \vec{\Pi}(t,r,\theta,\phi,\psi)
= \sum_{k\in K} e^{2ik\Omega t}\Phi_k (r) \vec{D}_k(\theta,\phi,\psi)\ .
\end{equation}
This has the eigenfrequencies $\omega=2k\Omega$ $(k\in K)$.  Since the solution has periodic time dependence on several frequencies, it is multi-oscillating.
In Ref.~\cite{Choptuik:2019zji}, multi-oscillating boson stars with non-commensurate frequencies have been constructed by solving partial differential equations. In this paper, resonating solutions are obtained by solving ordinary differential equations although the frequencies are commensurate.

One important difference in our solutions from those of \cite{Choptuik:2019zji} is the presence of a horizon.  In order for solutions to remain steady-state (i.e. independent of $\tau$), fields cannot pass through the horizon.  This restricts the frequency of the fields to be multiple of the angular frequency of the horizon, and hence any multi-oscillating solutions must have commensurate frequencies.  We expect non-commensurate multi-oscillating geons and boson stars to exist within this theory \eqref{Estnscalar}, but they would neither fall within our ansatz nor be the horizonless limit to a black hole.

\section{Phase diagram}
\label{sec:phase}

Finally, we put all the solutions together in a phase diagram in Fig.~\ref{phasediagram}.  We take $j=9/2$ in an ensemble with fixed-$(E,J)$.
We use $E-E_\textrm{geon}$ as the vertical axis for visibility.
\begin{itemize}
\item The extreme MPAdS is located on the black curve. Regular MPAdS black holes exist above this curve, while the MPAdS develops a naked singularity below it.
\item The red curve and line are for the gravitational black resonators and geons.
The curve on the top corresponds to the onset of the gravitational superradiant instability of the MPAdS.
The black resonators branch off from this curve to the bottom.
The horizontal red line in the bottom ($E=E_\textrm{geon}$) expresses the family of the gravitational geons.
The black resonators lie between the red curve and line.
\item The blue curves are associated with the hairy MPAdSs and boson stars.
The upper-left part of the blue curve, from $J=0$ to the black dot at $J=2.285$, is the onset of the scalar field superradiant instability on MPAdS black holes for $k=j=9/2$.
The onset coincides with the extreme MPAdS at the black dot, where the onset terminates.
The family of hairy MPAdS black holes branches off from this curve.

\begin{figure}
  \centering
\subfigure[Phase diagram]
 {\includegraphics[scale=0.5]{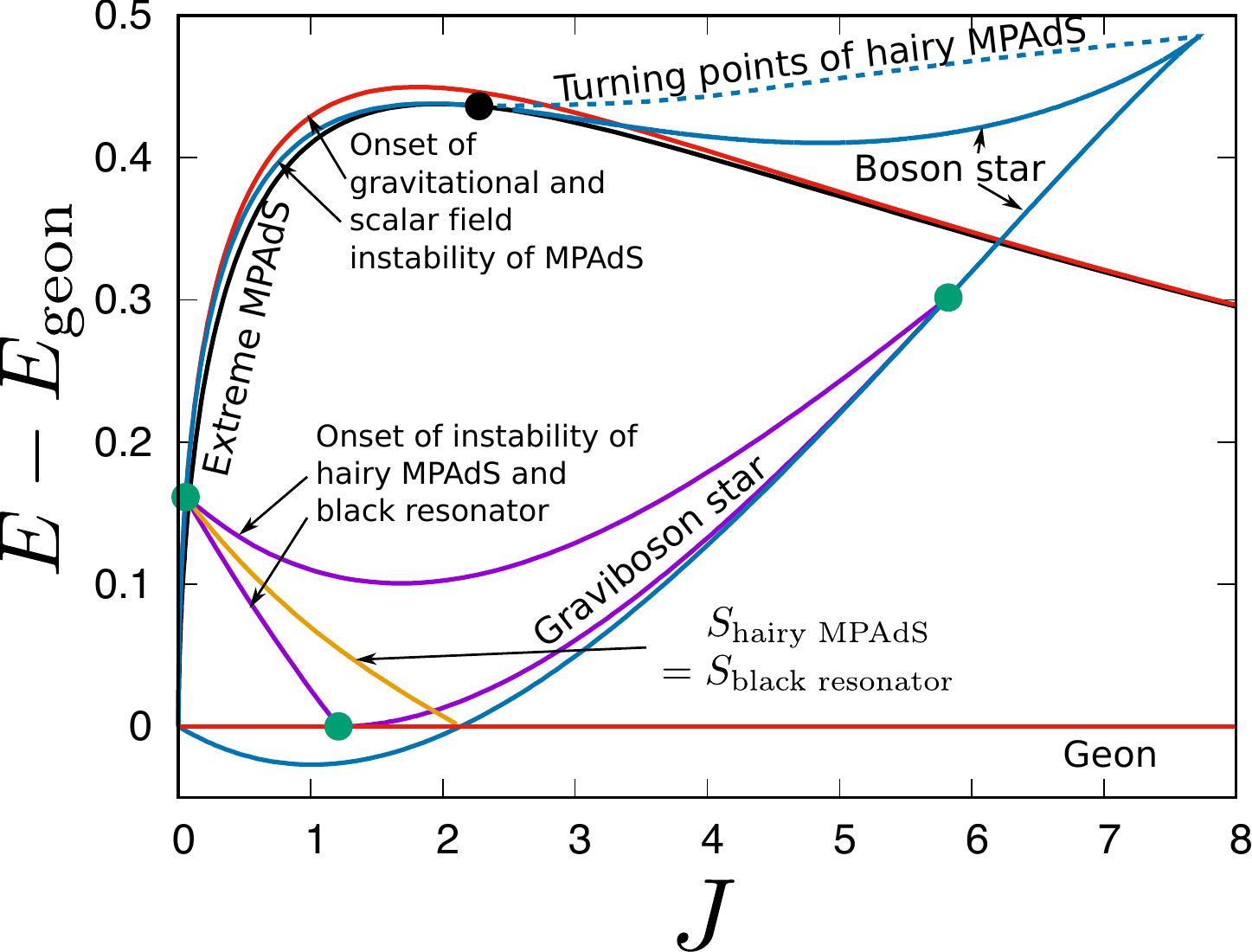}\label{phasediagram}
  }
  \subfigure[Black hole having maximal entropy]
 {\includegraphics[scale=0.5]{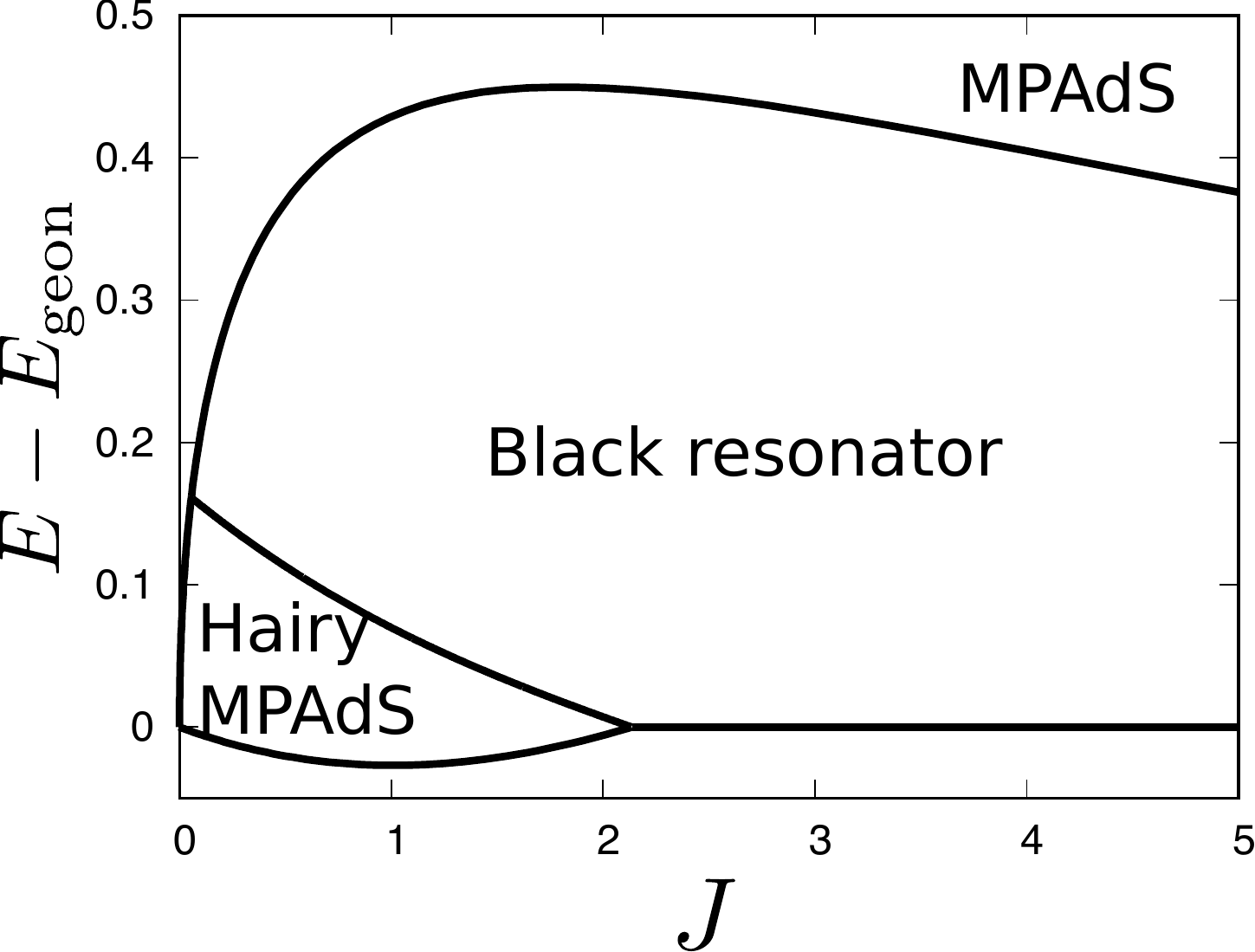}\label{phasebdry}
  }
 \caption{
(a) Phase diagram of asymptotically AdS solutions in Einstein-multiplet complex scalar fields system for $j=9/2$.
(b) Black hole having maximal entropy among the MPAdS, hairy MPAdS, black resonator, and hairy black resonator for $j=9/2$.
}
\end{figure}

The other blue curve corresponds to the family of boson stars.
There is a maximum in $(E,J)$ at the top right, which is a turning point for the curve for the boson star.
A collection of turning points for hairy MPAdSs, denoted by a blue dashed curve, extends from the top-right tip to the left as $r_h$ is increased.
It appears to extend toward the black dot.
The hairy MPAdSs exist in the region that is apparently enclosed by the blue curves for the onset and the boson star before the turning point, and the blue dashed curve.
In the upper-right region enclosed by the upper boson star onset curve and the dashed curve, physical quantities of hairy MPAdSs become multi-valued.
\item The purple curves denote the boundary of the existing region for the hairy black resonators and graviboson stars.
The bottom-right side of the distorted purple triangle curve is the locations of the family of graviboson stars.
The bottom-left side is the onset of the scalar field superradiant instability of the black resonator \cite{Ishii:2020muv}.  The top side is the onset of the gravitational instability of hairy MPAdS.
The hairy black resonators exist in the region enclosed by the purple curves.
\item The orange curve gives the location where the entropies of the black resonators and hairy MPAdSs become equal.
Across the transition, the entropy is continuous, but the field configurations are discontinuous. 
The hairy MPAdS has the higher entropy to the right of this curve, while the other side is dominated by the gravitational black resonators.
\end{itemize}
Fig.~\ref{phasebdry} is the phase diagram of the black hole solutions with the maximum entropy in the $(E,J)$-plane. The hairy black resonators never have the largest entropy and hence do not appear in this figure.

\section{Phase diagram for higher $j$}
\label{sec:phasej5}

We can also obtain solutions for the multiplet complex scalar with $j>9/2$. Here, we consider $j=5$.  Then, we need to consider $11$-component complex scalar fields at least.  Note that by setting one of the scalar multiplet components to zero, the $j=9/2$ solution is also a solution for the theory with $j=5$.  So though we have defined a scalar field ansatz for particular $j$'s, the solutions with different $j$ can be consistently compared to one another, so long as we choose the scalar field to have a sufficiently large multiplet.

In Ref.~\cite{Ishii:2020muv}, it was shown that, for an integer $j$, the scalar field is decomposed
into even and odd parity modes under the parity transformation
$P_1$ defined in Eq.(\ref{P1P20}). The even and odd parity modes satisfy $\Phi_{-k}= \Phi_k$ and $\Phi_{-k}=-\Phi_k$ $(k\in K)$, respectively.
In this section, we only consider the even parity mode. 

\begin{figure}
  \centering
\subfigure[Entropy for $j=9/2$ and $j=5$]
 {\includegraphics[scale=0.65]{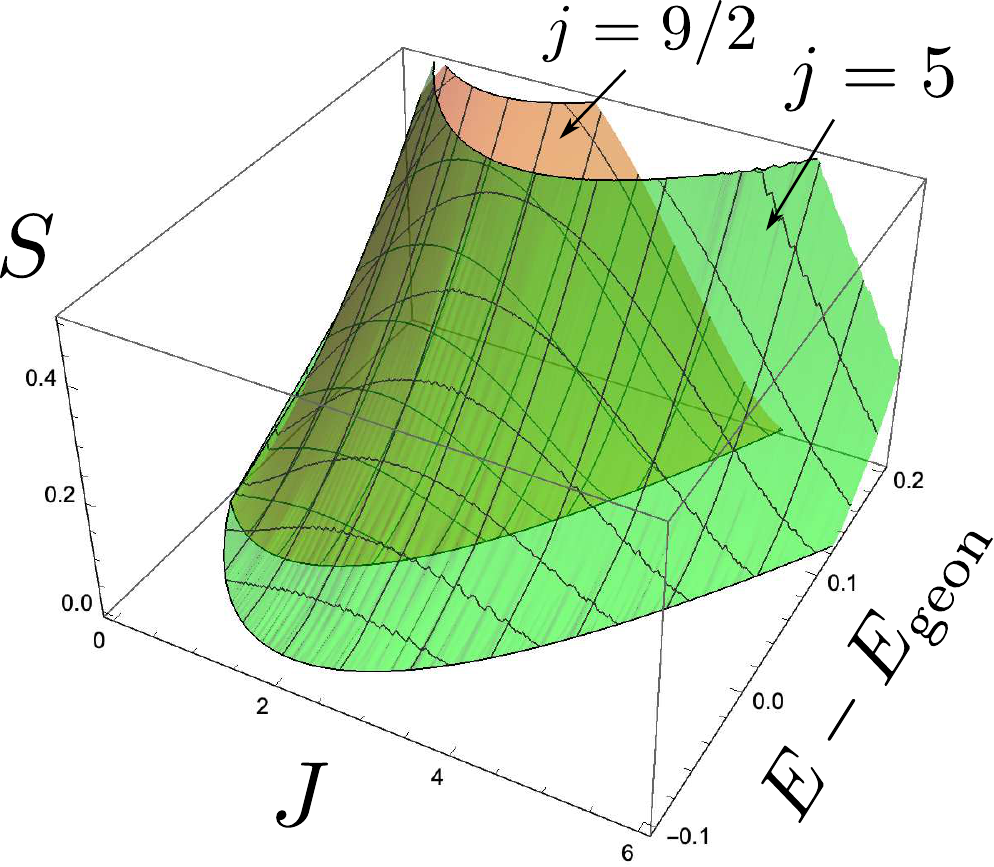}\label{SHMPAdS5}
  }
  \subfigure[Phase diagram for $j=5$]
 {\includegraphics[scale=0.5]{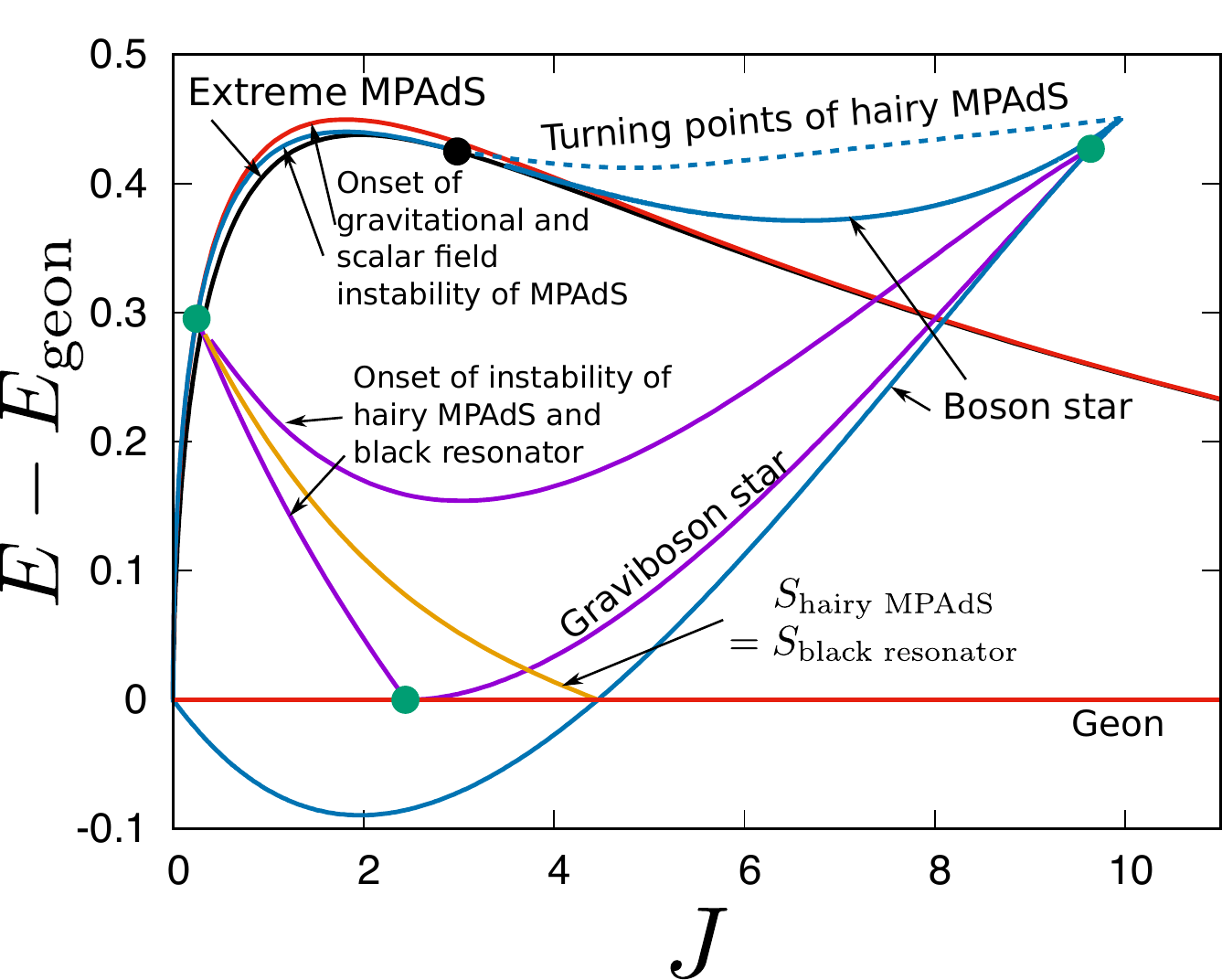}\label{phasediagram5}
  }
 \subfigure[Black hole having maximal entropy]
 {\includegraphics[scale=0.5]{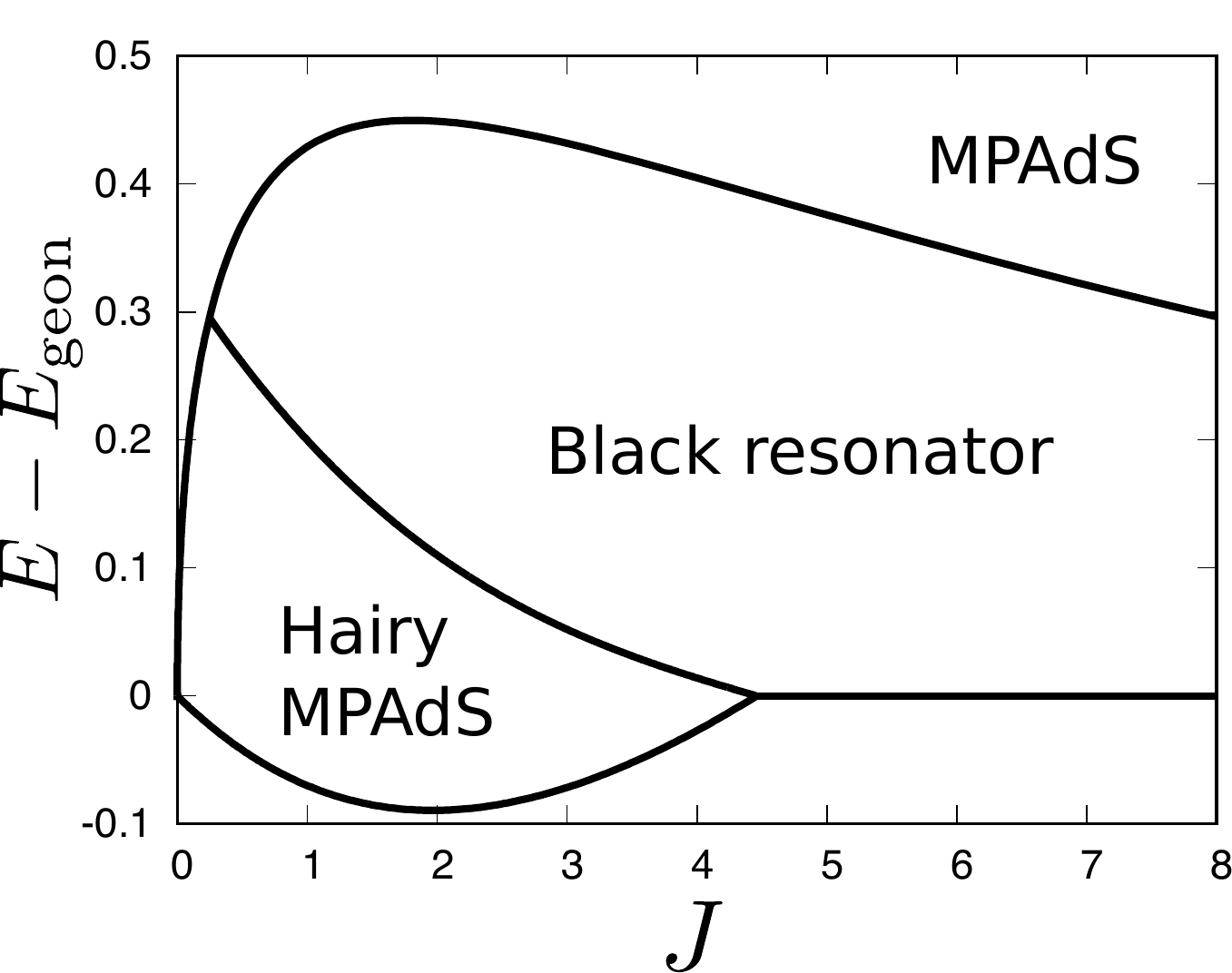}\label{phasebdry2j10}
  }
 \caption{
(a) Entropy of the hairy MPAdS for $j=9/2$ (orange) and $j=5$ (green).
(b) Phase diagram of asymptotically AdS solutions in Einstein-multiple complex scalar fields system for $j=5$.
(c) Black hole having maximal entropy among the MPAdS, hairy MPAdS, black resonator, and hairy black resonator for $j=5$.}
\end{figure}

In Fig.~\ref{SHMPAdS5}, we compare the entropies of the hairy MPAdS for $j=9/2$ and $j=5$.\footnote{
In the case of $\alpha(r)=1$, the equations of motion are identical for
$\Phi_{j}(r)=\cos \lambda\, \phi(r)$, $\Phi_{-j}(r)=\sin \lambda \, \phi(r)$
and $\Phi_k=0$ $(|k|\neq j)$ for any value of $\lambda$.
The even and odd parity modes correspond to $\lambda=\pm \pi/2$, and both modes give the same equations of motion.
}
We find that the solution for $j=5$ has higher entropy than that for $j=9/2$ at least in the region labeled as ``hairy MPAdS''
in Fig.~\ref{phasebdry}.

This result suggests that, in a theory with a $(2j'+1)$-component complex scalar field,
a hairy MPAdS with $j<j'$ evolves into that with $j=j'$ by the superradiant instability in the region of a small angular momentum
if we assume $SU(2)_L$ spacetime symmetry.
(Note that the hairy MPAdS with $j=j'$ should be further unstable to $SU(2)_L$-breaking perturbations~\cite{Green:2015kur}.)

Fig.~\ref{phasediagram5} is the phase diagram of solutions with $j=5$.
For the explanation of each curve, see section~\ref{sec:phase}.
The diagram is qualitatively similar to that for $j=9/2$.

Fig.~\ref{phasebdry2j10} is the phase diagram of the black hole solutions with the maximum entropy for $j=5$. 
The region in which the hairy MPAdS with $j=5$ entropically dominates is bigger compared to the case of $j=9/2$.
This indicates that, the larger the quantum number $j$ is, the wider the region covered by the hairy MPAdS will be in Fig.~\ref{phasebdry}.
If we extrapolate to arbitrarily large $j$, the black resonator would never dominate the phase diagram in a theory with an infinite number of complex scalar fields. 

\section{Conclusion}
\label{sec:conclude}

To summarise our results, we have studied asymptotically global AdS solutions of Einstein gravity coupled to a $(2j+1)$ complex scalar multiplet within a cohomogeneity-1 ansatz.  The following solutions are available within our ansatz: Myers-Perry-AdS black holes, black resonators (black holes with gravitational hair), black holes with scalar hair, and hairy black resonators (black holes with both gravitational and scalar hair).  The latter three of these branch from various superradiant instabilities and have zero horizon limits that are geons, boson stars, and graviboson stars, respectively. The phase diagram of all of these solutions was shown in Figs.~\ref{phasediagram} and \ref{phasediagram5} for $j=9/2$ and $j=5$, respectively.

The entropy of the hairy black resonator is never the largest among the three available solutions as shown in Fig.~\ref{phasebdry} for $j=9/2$. 
This seems natural in the view of the perturbative stability. Inside the triangular region enclosed by purple curves in Fig.~\ref{phasediagram}, both of the black resonator and hairy MPAdS is stable against corresponding perturbations.
Thus, the black resonator and hairy MPAdS would not evolve into the hairy black resonator. This is consistent with the fact that the hairy black resonator is entropically subdominant.

Finally, we were able to compare both $j=9/2$ and $j=5$ solutions, and we find that the $j=5$ hairy MPAdS solutions are dominant and cover a larger portion of phase space than those of $j=9/2$.  It is natural to expect that the trend continues to higher $j$. 
We can make this claim stronger by the following argument.
The phase boundary between hairy MPAdS and black resonators always lies between two points: (1) The intersection between gravitational and scalar onsets of MPAdS, and (2) where the boson stars intersect geons.
Results in Ref.~\cite{Ishii:2020muv} suggest that the point (1) is located at a higher angular momentum for a higher $j$. 
Also, by explicitly constructing boson stars for $j\leq 11/2$, we found that the same applies to the point (2) at least for $j=9/2,5,11/2$.

The study of time evolution of this system is an interesting future direction.
If we assume $SU(2)_L$-symmetry in the spacetime,
a time-dependent ansatz for this system would give a $1+1$ dimensional evolution system.
Curiously, though black resonators are unstable to superradiant scalar perturbations and hairy MPAdS are unstable to superradiant gravitational perturbations, the solutions that branch from these instabilities (namely, the hairy black resonators) are never entropically dominant.  Therefore, these unstable black resonators or hairy MPAdS cannot evolve to hairy black resonators.  Instead, unstable black resonators will likely evolve towards hairy MPAdS, removing its gravitational hair.  Similarly, unstable hairy MPAdS will evolve towards black resonators, shedding its scalar hair.
Hairy black resonators themselves can (entropically) evolve to either black resonators or hairy MPAdS, 
most likely to whichever is dominant. In all of these case, scalars with $k<j$ will be suppressed, and either the gravitational or $k=j$ scalar field instability will survive.

It would be especially interesting to study the time evolution of a system with large $j$.  As we have mentioned, a large $j$ multiplet contains smaller $j$ solutions within it, so the full system contains a tower of lower wavenumbers, many of which are unstable modes in black resonators or MPAdS.  These modes have different growth rates, with the largest wavenumber typically being the slowest.  However, the hairy black hole with the largest wavenumber is likely the most dominant entropically.  A time evolution would therefore tell us how these competing instabilities interact with each other, and how a cascade to higher wavenumbers proceeds.

Though the growth rates of high modes are extremely small and present a significant numerical challenge, such a calculation seems more feasible in this 1+1 setting than the full 3+1 setting of Kerr-AdS.  Furthermore, the high-wavenumbers are associated with angular directions rather than the radial direction.  As the 1+1 equations do not directly see the angular gradients, the numerical resolution can be kept relatively low, allowing faster time evolution due to the larger Courant number.

A time-dependent ansatz for this system would give a $1+1$ dimensional evolution system. In the discrete isometries~(\ref{P1P2}), we can only assume $P_2$-invariance for the time-dependent spacetime, and the equations of motion require a non-trivial cross term like $\gamma(t,r)\sigma_1\sigma_2$ in the metric for consistency.

Despite the fact that we can construct solutions with several values of $j$, this comes at a cost. In particular, we would like to remind the reader that for each value of $j$ we need $2j+1$ complex scalar fields to make our co-homogeneity one ansatz work. Furthermore, the phases of each of these scalars must be fine tuned so that the overall dependence in the angles do cancel. Perhaps more importantly, all these scalars are minimally coupled to gravity.\footnote{One could potentially add a mass term, and much of our discussion would still go through.} One might ask whether such scalars are easy to come by in consistent reductions from some higher dimensional supergravity theory such as type IIB and the answer appears to be no. To our knowledge the largest known conjectured truncation of type IIB supergravity\footnote{This has actually never been shown in full generality, partially because of the self dual condition imposed
on the Ramond-Ramond F5 form flux, even though interesting progress has been recently made in \cite{Ciceri:2014wya}.} arises when considering compactifications of the form AdS$_5\times S^5$, with the lower dimensional theory being five-dimensional $\mathcal{N}=8$ gauged supergravity comprising a total 42 scalars, 15 gauge fields and 12 form fields. However, these scalars appear to be non-minimally coupled to gravity, and to have very complicated potentials (see for instance \cite{Ciceri:2014wya}), thus giving very little hope that our model will find a precise holographic realisation.

\acknowledgments
We would like to thank Oscar~Dias for useful conversations.
The work of T.~I.~was supported in part by JSPS KAKENHI Grant Number JP18H01214 and JP19K03871.
The work of K.~M.~was supported in part by JSPS KAKENHI Grant Number JP18H01214 and JP20K03976.
BW acknowledges support from ERC Advanced Grant GravBHs-692951 and MEC grant FPA2016-76005-C2-2-P. J.~E.~S. is supported in part by STFC grants PHY-1504541 and ST/P000681/1. J.~E.~S. also acknowledges partial support from a J. Robert Oppenheimer Visiting Professorship.

\appendix

\section{Technical details}
\label{tech}
In this appendix, we collect technical details for solving the Einstein-complex scalar multiplet system.

\subsection{Equations of motion}

With our ansatz, the equations of motion are given by coupled ODEs.
From Eqs.~(\ref{SU2metric}) and (\ref{Tab}), the trace $\mathcal{T}$ defined in Eq.~(\ref{Tdef}) can be computed as
\begin{multline}
 \mathcal{T}=\sum_{k\in K}\bigg[(1+r^2)g\Phi'_{k}{}^2
+\frac{2}{r^2}\left(\alpha-\frac{1}{\alpha}\right)\epsilon_{k-1}\epsilon_k \Phi_{k-2}\Phi_k\\
+\frac{1}{r^2}\left(\alpha+\frac{1}{\alpha}\right)(\epsilon_{k}^2+\epsilon_{k+1}^2)\Phi_k^2
+4\left(\frac{1}{r^2\beta}-\frac{h^2}{(1+r^2)f}\right) k^2 \Phi_k^2
\bigg]\ .
\label{trT}
\end{multline}
For the metric ansatz (\ref{SU2metric}), the Einstein equations $G_{\mu\nu}-6g_{\mu\nu}=T_{\mu\nu}$ become
\begin{align}
\begin{split}
f'=&\frac{1}{r(1+r^2)^2 g \alpha^2 (r\beta'+6\beta)}
[
4 r^2 h^2 (\alpha^2-1)^2 \beta\\
&+r (r^2+1) g\{r(1+r^2)  f \alpha'{}^2 \beta
-r^3   h'{}^2 \alpha^2 \beta^2
-2 (2+3 r^2)  f \alpha^2   \beta'\}\\
&
-4 (1+r^2) f  \{6 r^2 \alpha^2 \beta (g-1)+3 g \alpha^2 \beta + (\alpha^2-\alpha \beta+1)^2-4 \alpha^2\}
]\\
&+\frac{4rf\beta}{r\beta'+6\beta}T_{rr}
\label{EOMf}
\end{split}
\ ,\\
\begin{split}
g'=&\frac{1}{6 r(1+r^2)^2 f \alpha^2 \beta}
[
-4 r^2 h^2   (\alpha^2-1)^2\beta\\
&+r(1+r^2)g\{
-r (1+r^2)  f \alpha'{}^2 \beta
+r^3 h'{}^2 \alpha^2 \beta^2\\
&- (-r(1+r^2) f'+2 f) \alpha^2 \beta'\}
+4 (1+r^2) f  \{-6r^2\alpha^2 \beta(g - 1)-3 g \alpha^2 \beta \\
& \hspace{28ex}
+\alpha^4+4 \alpha^3 \beta-5 \alpha^2 \beta^2-2 \alpha^2+4 \alpha \beta+1\}
]\\
&-\frac{2}{3r(1+r^2)^2f\beta}[
r^2\beta T_{tt}
-4 r^2 h\beta T_{t3}
+4(r^2h^2 \beta  - (1+r^2)f)T_{33}]\\
&+\frac{1}{3r(1+r^2)}\left[\left(\alpha-\frac{1}{\alpha}\right)(T_{++}+T_{--})
-2\left(\alpha + \frac{1}{\alpha}\right)T_{+-}\right]
\label{EOMg}
\end{split}
\ ,\\
\begin{split}
h''=&\frac{1}{2 r^2 (1+r^2) \alpha^2 \beta f g}
[
8 f h (\alpha^2-1)^2\\
&-r (1+r^2) h' \alpha^2 \{r (f g' \beta -f' g \beta +3 f g \beta')+10 f g \beta \}
]+\frac{4(2 h T_{33}  - T_{t3})}{r^2 (1+r^2) g \beta}
\label{EOMh}
\end{split}
\ ,\\
\begin{split}
\alpha''=&\frac{1}{2 r^2 (1+r^2)^2 f \alpha g \beta}
[
2 r^2 (r^2+1)^2  f g \alpha'{}^2 \beta\\
&-r (r^2+1) \alpha \alpha'  \{r(1+r^2)(f g \beta)' +2 (3+5 r^2)  f g \beta\}\\
&-8 (\alpha^2-1) \{
r^2 h^2 \beta (\alpha^2 + 1)-(1+r^2) f \alpha(\alpha-\beta) -(1+r^2)f
\}]\\
&-\frac{\alpha}{r^2(1+r^2) g}\left[\left(\alpha + \frac{1}{\alpha}\right)(T_{++}+T_{--})
-2\left(\alpha - \frac{1}{\alpha}\right)T_{+-}\right]
\label{EOMa}
\end{split}
\ ,\\
\begin{split}
\beta''=&\frac{1}{(2 r^2 (1+r^2)) f g \alpha^2 \beta}
[
-2r^4  g h'{}^2 \alpha^2 \beta^3 \\
&-r \alpha^2 \beta' \{
r(1+r^2)(f' g \beta+ f g' \beta- f g \beta')
+2 (3+5 r^2)f g \beta \}\\
&-8 f \beta (\alpha^4+\alpha^3 \beta-2 \alpha^2 \beta^2-2 \alpha^2+\alpha \beta+1)
]-\frac{8}{r^2 (1+r^2)}T_{33}\\
&-\frac{\beta}{r^2(1+r^2) g}\left[\left(\alpha-\frac{1}{\alpha}\right)(T_{++}+T_{--})
-2\left(\alpha + \frac{1}{\alpha}\right)T_{+-}\right]
\label{EOMb}
\end{split}
\ .
\end{align}
For our scalar field ansatz (\ref{PiSU2}), the Klein-Gordon equation $\Box \, \vec{\Pi}=0$ gives $|K|=\lfloor j+1 \rfloor$ equations of the form
\begin{equation}
L_k \Phi_k+c_{k-1} \Phi_{k-2} +c_{k+1} \Phi_{k+2} =0\ ,
\label{phieq2}
\end{equation}
where
\begin{multline}
L_k=(1+r^2)g \frac{\mathrm{d}^2}{\mathrm{d}r^2}
 +\left[\frac{1+r^2}{2}\left(
\frac{f'}{f}+\frac{g'}{g}+\frac{\beta'}{\beta}
\right)+\frac{3+5r^2}{r}\right]g\frac{\mathrm{d}}{\mathrm{d}r}\\
-\frac{\epsilon_k^2+\epsilon_{k+1}^2}{r^2}\left(\alpha+\frac{1}{\alpha}\right)-\frac{4k^2}{r^2\beta}
+\frac{4k^2h^2}{(1+r^2)f}\ ,
\end{multline}
and
\begin{equation}
c_k=-\frac{\epsilon_{k}\epsilon_{k+1}}{r^2}\left(\alpha-\frac{1}{\alpha}\right)\ .
\end{equation}
In (\ref{phieq2}), the mode coupling is ``double-stepping'' --- the mode with $k$ is coupled to those with $k\pm 2$.

We integrate these equations from the horizon $r=r_h$ (or the center $r=0$ if the geometry is horizonless) to infinity $r \to \infty$ by using the 4th order Runge-Kutta method.
The boundary conditions for solving the equations are given below.

\subsection{Boundary conditions at infinity}
\label{app:bc_inf}

We require the spacetime to be asymptotically AdS at infinity. In the rotating frame at infinity, the condition for the metric components is
\begin{equation}
f,\alpha,\beta \to 1, \quad h \to \Omega \quad (r\to\infty)\ ,
\label{asympAdS}
\end{equation}
and then $g\to 1$ also follows from the other the equations of motion.
The asymptotic value of $h(r)$ actually corresponds to the angular velocity of the horizon $\Omega$ as explained in section~\ref{sec:scalar_coh1}.
We also require that the massless scalar field falls off at infinity,
\begin{equation}
 \Phi_k \to 0 \quad(r\to\infty)\ .
\label{Psiinf}
\end{equation}
In the interpretation of the gauge/gravity duality, this means that there is no external source for the dual scalar operators in the boundary field theory.
Thus, nontrivial scalar fields are spontaneously induced by the instability in the geometry.

However, imposing $f \to 1$ as $r \to \infty$ is actually redundant because there is rescaling symmetry in Eq.~(\ref{SU2metric}).
By a coordinate transformation $\tau\to c \, \tau$ for a constant $c$, the line element is invariant if the metric components are rescaled as
\begin{equation}
 f(r)\to \frac{f(r)}{c^2}\ ,\quad
 h(r)\to \frac{h(r)}{c}\ .
\label{fhrescale}
\end{equation}
Hence, if we obtain a solution with $f_\infty\equiv f(r=\infty) \neq 1$, we can rescale it so that the new $f$ satisfies $f\to 1$ at infinity.
That is, we take $c=\sqrt{f_\infty}$ in the above scaling equation.
Therefore, when we solve the equations of motion, we only need to impose
\begin{equation}
\alpha,\beta \to 1\quad(r\to\infty)\ ,
\label{asympAdS2}
\end{equation}
and then we can apply the transformation~(\ref{fhrescale}) to obtain a rescaled solution satisfying $f\to 1$.

\subsection{Boundary conditions at the origin for horizonless solutions}
\label{app:bc_horless}

For horizonless solutions, we require regularity at the origin of AdS $r=0$.
To avoid a conical singularity at $r=0$, we impose
\begin{equation}
 g,\alpha,\beta \to 1\quad (r\to 0)\ .
 \label{regularity_origin}
\end{equation}
Then, from Eq.~(\ref{phieq2}), the regular solution of the scalar field near $r=0$ has the behavior
\begin{equation}
 \Phi_k \sim r^{2j}\ .
\end{equation}
To handle this behavior at arbitrary $j$, we find it convenient to redefine the scalar field as
\begin{equation}
 \Phi_k(r) = \left(\frac{r^2}{1+r^2}\right)^j \Psi_k(r)
\end{equation}
and solve the equations of motion for new variables $X \equiv (f,g,h,\alpha,\beta,\Psi_k)$.
They can be expanded near $r=0$ as
\begin{equation}
 X(r)=\sum_{m=0}^\infty X_{2m}r^{2m}\ ,
\end{equation}
where $g_0=\alpha_0=\beta_0=1$ as Eq.~(\ref{regularity_origin}).
Substituting this expansion into Eqs.~(\ref{EOMf}-\ref{phieq2}) and specifying $f_0, h_0, \alpha_2, \beta_2, \Psi_{k,0}$ as the input parameters, we can determine the higher order coefficients $X_{2m}$ order by order.
(In practice, we evaluated $X_{2m}$ for $m\leq 2$.)
Because of the rescaling (\ref{fhrescale}), we can set $f_0=1$ without loss of generality.

To construct gravitational geons, we set the scalar field zero, $\Psi_k(r)=0$ for all $k$.
Then, we have three free parameters to be specified at the origin, $h_0, \alpha_2$, and $\beta_2$, while there are two boundary conditions~(\ref{asympAdS2}) at infinity.
Thus, the geons are obtained as a one-parameter family.
We set $\alpha_2\neq 0$ as the input parameter and determine $h_0$ and $\beta_2$ by the shooting method so that Eq.~(\ref{asympAdS2}) is satisfied.
To obtain a family of geons, we start from a normal mode of pure AdS where $h_0=3/2$ and $\beta_2=0$ and turn on a tiny $\alpha_2$.
Once a solution is successfully obtained, we slightly vary the value of $\alpha_2$ as well as using the previous result as the initial guess for the next solution.
We repeat this process and construct the geon solutions shown in Fig.~\ref{Egeon}.

For boson stars, we set $\alpha(r)=1$ and $\Psi_k(r)=0$ for $k<j$, which means that $\alpha_2=\Psi_{k<j,0}=0$.
There are three free parameters $h_0, \beta_2, \Psi_{j,0}$ at the origin.
At infinity, we impose two conditions $\beta\to 1$ and $\Psi_j\to 1$.
Similar to the case of geons, we set $\Psi_{j,0} \neq 0$ as the input and start the shooting method from the initial guess given by the scalar field normal mode in pure AdS with $h_0=1+2/j$ and $\beta_2=0$.

For the construction of graviboson stars, we have $|K|+3$ free parameters at the origin $h_0, \alpha_2, \beta_2, \Psi_{k,0}$, and $|K|+2$ conditions~(\ref{Psiinf}) and (\ref{asympAdS2}) at infinity.
We set $\Psi_{j,0}\neq 0$ as the input and determine the other $|K|+2$ parameters by the shooting.
The normal mode frequencies $\omega$ for scalar fields in gravitational geon backgrounds have been computed in Ref.~\cite{Ishii:2020muv}.
The onset for graviboson stars corresponds to $\omega=0$.
We start from that point and turn on a tiny value of $\Psi_{j,0}$.

\subsection{Boundary conditions at the horizon}
\label{app:bc}

For black hole solutions, boundary conditions are imposed at the horizon $r=r_h$.
The field variables $Y\equiv (f,g,h,\alpha,\beta,\Phi_k)$ can be expanded near the horizon as
\begin{equation}
 Y(r)=\sum_{m=0}^\infty Y_m (r-r_h)^m\ ,
\label{Yepand}
\end{equation}
where $f_0=g_0=0$.
Substituting Eq.~(\ref{Yepand}) into the equation of motion for $h(r)$ (\ref{EOMh}) and looking at the leading order, we obtain
\begin{equation}
\left\{(\alpha_0-1)^2+2\alpha_0^2\sum_{k\in K}k^2 \Phi_{k,0}^2\right\}h_0=0\ .
\end{equation}
There are two possibilities for the solutions to this equation:
(i) $\alpha_0=1$ and $\Phi_{k,0}=0$,
and
(ii) $h_0=0$.
The case (i) is nothing but the MPAdS solution. Indeed, the horizon value $h_0$ can be arbitrary because of the recovered $U(1)_R$ isometry.
For hairy and resonating solutions, we consider (ii).
Let us redefine the field variables as
\begin{equation}
 f(r)=F(r) \tilde{f}(r)\ ,\quad
 g(r)=F(r) \tilde{g}(r)\ ,\quad
 h(r)=F(r) \tilde{h}(r)\ ,
\end{equation}
where $F(r)\equiv 1-r_h^6/r^6$.
With the new variables, the near horizon expansion can be given by
\begin{equation}
 Z(r)=\sum_{m=0}^\infty Z_m (r-r_h)^m\ ,
\label{Zepand}
\end{equation}
where $Z=(\tilde{f},\tilde{g},\tilde{h},\alpha,\beta,\Phi_k)$.
This $F(r)$ is chosen by hand so as to accommodate the near horizon behavior and employ $\tilde{f}_0,\tilde{g}_0,\tilde{h}_0 \neq 0$.\footnote{The choice of $F(r)$ can be arbitrary, but we assume $F(r_h)=0$ and $F'(r_h)\neq0$.}
Substituting Eq.~(\ref{Zepand}) into Eqs.~(\ref{EOMf}-\ref{phieq2}) and taking $\tilde{f}_0,\tilde{h}_0,\alpha_0,\beta_0,\Phi_{k,0},r_h$ as the input parameters, we can determine the higher order coefficients $Z_m$.
(Practically in our calculations, we truncate the series to $m\leq 2$.)
Furthermore, we can set $\tilde{f}_0=1$ without loss of generality because of the rescaling symmetry \eqref{fhrescale}.

For black resonators, the scalar fields are trivial, $\Phi_k(r)=0$. There are four free parameters
$\tilde{h}_0,\alpha_0,\beta_0,r_h$ at the horizon and two conditions~(\ref{asympAdS2}) at infinity.
Thus we need to specify two input parameters at the horizon, for which we choose $\alpha_0$ and $r_h$, and then $\tilde{h}_0$ and $\beta_0$ are determined by the shooting method.
We start from the onset of the superradiant instability of the MPAdS evaluated in \cite{Murata:2008xr,Ishii:2020muv}, where $\alpha_0=1$,
and turn on a small deformation as $\alpha_0-1 \neq 0$.
Once the shooting method converges and a black resonator solution is obtained, we slightly increase $\alpha_0-1$ as well as using the previous data as the initial guess of the shooting method for the next solution.

For hairy MPAdS, we set $\alpha(r)=1$ and $\Phi_k(r)=0$ for $k<j$. Therefore, we have $\alpha_0=1$ and $\Phi_{k<j,0}=0$ at the horizon.
We are left with four free parameters $\tilde{h}_0,\beta_0,\Phi_{j,0},r_h$ at the horizon, and there are two conditions $\beta\to 1$ and $\Phi_j\to 0$ at infinity.
We specify $\Phi_{j,0}$ and $r_h$ as input parameters and determine $\tilde{h}_0$ and $\beta_0$ by matching the boundary condition at infinity.
The scalar field superradiant instability of the MPAdS \cite{Cardoso:2013pza,Ishii:2020muv} is where the hairy solutions branch.

For hairy black resonators, there are $|K|+4$ free parameters $\tilde{h}_0,\alpha_0,\beta_0,\Phi_{k,0},r_h$ at the horizon and $|K|+2$ conditions~(\ref{Psiinf}) and (\ref{asympAdS2}) at infinity.
There are two options in extending hairy black resonator solutions: from the scalar field superradiant instability of black resonators, and from the gravitational one of hairy MPAdSs.
We take the latter.
The gravitational instability of the hairy MPAdS is studied in appendix~\ref{InstHMPAdS}.
We start the construction of the hairy black resonators from the onset of this instability.
We specify $\alpha_0$ and $r_h$ as the input parameters and determine the others by the shooting method.
Starting from the onset of the gravitational instability of the hairy MPAdS, we turn on a small $\alpha_0-1$.
We could also obtain the solutions branching off from the scalar field superradiant instability of black resonators evaluated in~\cite{Ishii:2020muv}.
However, it turned out that the shooting method did not converge nicely around the onset of the instability, and therefore we resort to the other option.
This problem is discussed in appendix~\ref{fitting}.

\subsection{Physical quantities}

Near the asymptotic infinity $r\to \infty$, the asymptotic solutions of the metric components and scalar fields are
\begin{equation}
\begin{split}
f(r)&=1+\frac{c_f}{r^4}+\cdots\ ,\quad
g(r)=1+\frac{c_f+c_\beta}{r^4}+\cdots\ ,\quad
h(r)=\Omega+\frac{c_h}{r^4}+\cdots\ ,\\
\alpha(r)&=1+\frac{c_\alpha}{r^4}+\cdots\ ,\quad
\beta(r)=1+\frac{c_\beta}{r^4}+\cdots\ ,\quad
\Phi_k(r)=\frac{c_k}{r^4}+\cdots\ ,
\end{split}
\label{asym}
\end{equation}
where $c_f, c_h, c_\alpha, c_\beta$, and $c_k$ are the constants that are determined by matching the series with the bulk, and the source of $\Phi_k$ has already been set zero.
As discussed in section~\ref{sec:scalar_coh1},
the asymptotic value of $h(r)$ corresponds to the angular velocity $\Omega$.
The constant $c_k$ corresponds to the expectation value of the operator $\vec{\mathcal{O}}$ dual to $\vec{\Pi}$ in the boundary theory as
\begin{equation}
 \langle \vec{\mathcal{O}} \rangle
= \sum_{k\in K} c_k \vec{D}_k (\theta,\phi,\chi)
= \sum_{k\in K} c_k e^{-2k\Omega t} \vec{D}_k (\theta,\phi,\psi)\ .
\end{equation}
This depends on both time and spatial coordinates on the boundary.

Because the boundary source for the scalar field is absent, the boundary energy-momentum tensor $T_{ij}$ can be given by \cite{Ashtekar:1999jx,Balasubramanian:1999re,deHaro:2000vlm,Kinoshita:2008dq}.
\begin{equation}
 8\pi G_5 T_{ij}=-\frac{r^2}{2}C_{i\rho j \sigma} n^\rho n^\sigma\bigg|_{r=\infty}\ ,
 \label{TijFromWeyl}
\end{equation}
where $i$ and $j$ run over the coordinates on the AdS boundary,
$n^\mu$ is the unit normal to a bulk $r$-constant surface,
and $C_{\mu\nu\rho\sigma}$ is the bulk Weyl tensor.
Using Eq.~(\ref{asym}), we obtain
\begin{multline}
 8\pi G_5 T_{ij}\mathrm{d}x^i \mathrm{d}x^j
= \frac{1}{2}(c_\beta-3c_f) \mathrm{d}\tau^2
+2 c_h \mathrm{d}\tau(\sigma_3+2\Omega \mathrm{d}\tau)
-\frac{c_f+c_\beta}{8}(\sigma_1^2+\sigma_2^2)\\
+\frac{c_\alpha}{2}(\sigma_1^2-\sigma_2^2)
+\frac{1}{8}(-c_f+3c_\beta)(\sigma_3+2\Omega \mathrm{d}\tau)^2
\ .
\end{multline}
This is written in the rotating frame at infinity $(\tau,\chi)$.
In the non-rotating frame $(t,\psi)$, the boundary stress tensor is rewritten as
\begin{multline}
 8\pi G_5 T_{ij}\mathrm{d}x^i \mathrm{d}x^j
= \frac{1}{2}(c_\beta-3c_f) \mathrm{d}t^2
+2 c_h \mathrm{d}t \bar{\sigma}_3-\frac{c_f+c_\beta}{8}(\bar{\sigma}_1^2+\bar{\sigma}_2^2)\\
+c_\alpha (e^{4i\Omega t}\bar{\sigma}_+^2+e^{-4i\Omega t}\bar{\sigma}_-^2)
+\frac{1}{8}(-c_f+3c_\beta)\bar{\sigma}_3^2
\ ,
\label{Tmunu_non-rot}
\end{multline}
where $\bar{\sigma}_i \ (i=1,2,3)$ are the invariant one-forms defined in the non-rotating frame:
$\chi$ in \eqref{inv1form} is replaced with $\psi$ for $\bar{\sigma}_i$.
The energy and angular momentum are given by
\begin{equation}
E=\int \mathrm{d}\Omega_3 T_{tt} =\frac{\pi(c_\beta-3c_f)}{8 G_5}\ ,\quad
J=-\int \mathrm{d}\Omega_3 T_{t (\psi/2)} =-\frac{\pi c_h}{2 G_5}\ ,
\label{EJdef}
\end{equation}
where we define the angular momentum with respect to $\psi/2\in [0,2\pi)$.

For black hole solutions,
the entropy $S$ and temperature $T$ are given by
\begin{equation}
S=\frac{\pi^2 r_h^3 \sqrt{\beta(r_h)}}{2G_5}\ ,\quad
T=\frac{(1+r_h^2)\sqrt{f'(r_h)g'(r_h)}}{4\pi}\ .
\end{equation}
For simplicity, we set $G_5=1$ in this paper.
We can easily recover the dependence on $G_5$ by
$E\to G_5 E$, $J\to G_5 J$, and $S\to G_5 S$.

\subsection{Interpolation of hairy black resonator data}
\label{fitting}

\begin{figure}
\begin{center}
\includegraphics[scale=0.6]{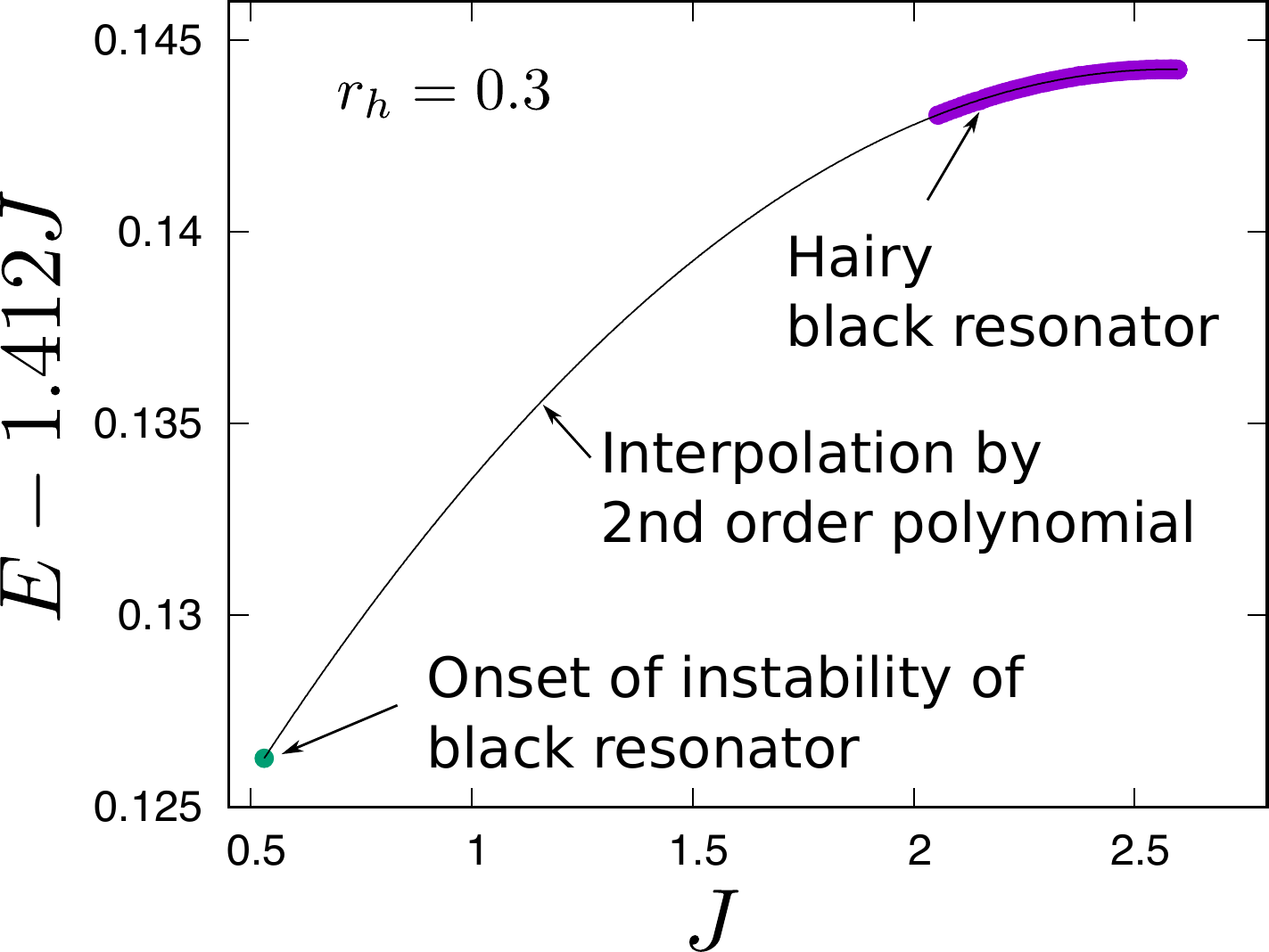}
\end{center}
\caption{
Energy of the hairy black resonator as a function of the angular momentum $J$ for a fixed horizon radius $r_h=0.3$.
}
\label{HBRinpol}
\end{figure}

When we construct the hairy black resonator, we take the route to extend the solution from the onset of instability of the hairy MPAdS.
Near the onset, the shooting method converges well, and we were able to obtain hairy black resonators.
As the solution approaches the onset of the instability of the black resonator, however, we find that the shooting method fails to converge.
Here we argue that the void, however, can be filled by interpolation.
In Fig.~\ref{HBRinpol}, we show $E$ of the hairy black resonator as a function of $J$ for a fixed horizon radius $r_h=0.3$.
For visibility, we use $E-1.412J$ as the vertical axis.
The purple points denote the numerical data, and the green point is the onset of the scalar field instability of the black resonator.
Between these points, we were not able to obtain numerical solutions.
There might be other solutions with similar parameters at the horizon.
One of the candidates is the black resonator with the scalar hair with nontrivial radial overtones.
Nevertheless, our data is fitted well by a second order polynomial as shown in the black curve in the figure.
We used the interpolated data in the corresponding region in Fig.~\ref{S3sol}.

\section{Instability of hairy MPAdS}
\label{InstHMPAdS}

In this appendix, we consider the linear perturbation of the hairy MPAdS.
To find the onset of the instability, we only need to consider $\tau$-independent perturbations and find their normal modes.
For the hairy MPAdS solutions, we have $\alpha(r)=1$ and $\Phi_k(r)=0$ ($k< j$), while the other functions have nontrivial $r$-dependence.
Around this background, we perturb the variables as $(f,g,h,\alpha,\beta,\Phi_k)\to (f+\delta f,g+\delta g,h+\delta h,\alpha+\delta \alpha,\beta+\delta \beta,\Phi_k+\delta\Phi_k)$ in Eqs.~(\ref{EOMf}-\ref{phieq2}) and keep the linear order in the perturbations.
It turns out that the perturbation variables $\delta \alpha$ and $\delta \Phi_{j-2}$ decouple from the others.
Their perturbation equations are given by
\begin{align}
\begin{split}
\delta\alpha''=&
- \frac{1}{2}\left\{\frac{(f g \beta)'}{f g \beta} +\frac{2 (3+5 r^2)}{r (1+r^2)}\right\}\delta\alpha'  \\
& -\frac{8}{(1+r^2)g}\left\{
\frac{\beta-2}{r^2\beta}+\frac{2h^2}{(1+r^2)f}
-\frac{T_{+-}}{4r^2}\right\}\delta\alpha
-\frac{\delta T_{++}+\delta T_{--}}{r^2(1+r^2) g}
\end{split}\ , \label{daeq} \\
L_{j-2} \delta \Phi_{j-2}=& \ \frac{4\sqrt{j(2j-1)}}{r^2} \Phi_j
\delta \alpha \ , \label{dpsieq}
\end{align}
where
\begin{align}
 T_{+-}=&-\frac{r^2}{4}\left[(1+r^2)g\Phi'_j{}^2
+4j^2\left(\frac{1}{r^2\beta}-\frac{h^2}{(1+r^2)f}\right) \Phi_j^2\right]\ , \\
\begin{split}
 \delta T_{\pm\pm}
=&-2\sqrt{j(2j-1)} \Phi_j \delta\Phi_{j-2}\\
&-\frac{r^2}{4}\left[
(1+r^2)g\Phi'_j{}^2
+\frac{4j}{r^2}\Phi_j^2
+4j^2\left(\frac{1}{r^2\beta}-\frac{h^2}{(1+r^2)f}\right) \Phi_j^2
\right]\delta \alpha
\end{split}\ .
\end{align}

We wish to find normal mode solutions to these equations with the sourceless boundary condition at infinity: $\delta \alpha\to 0$ and $\delta\Phi_{j-2}\to 0$.
On the black hole horizon, we require regularity.
Substituting $\delta \alpha=a_0 + a_1(r-r_h)+\cdots$ and $\delta \Phi_{j-2}=\phi_0 + \phi_1(r-r_h)+\cdots$ into Eqs.~(\ref{daeq}-\ref{dpsieq}), we can obtain the regular series solution where $a_i$ and $\phi_i$ for $i\geq1$ are determined by $a_0$ and $\phi_0$.
We set $a_0=1$ to fix the scale of the linear perturbation.
Then, $\phi_0$ is left as the free parameter on the horizon.
This parameter is tuned so that the boundary condition at infinity, $\delta \Phi_{j-2} \to 0$, is satisfied.
For a general hairy MPAdS background specified by the horizon value of the background scalar field $\Phi_j|_{r=r_h}$,
however, $\delta \alpha \to 0$ is not realized at infinity.
The boundary condition for $\delta \alpha$ is satisfied only at special values of $\Phi_j|_{r=r_h}$.
We monitor the value of $\delta \alpha|_{r=\infty}$ by increasing $\Phi_j|_{r=r_h}$ and search the value of $\Phi_j|_{r=r_h}$ at which $\delta \alpha|_{r=\infty}=0$ is satisfied.
We repeat this process for each horizon radius $r_h$.
For $j=9/2$, the result of the onset is the upper purple curve in Fig.~\ref{trig}.
From Fig.~\ref{onset_Om},
we know that MPAdS at the branching points of hairy MPAdS becomes unstable against gravitational perturbation
for $r_h\gtrsim 0.37$. This indicate that, in the upper side of the onset curve in Fig.~\ref{trig}, hairy MPAdS is unstable.

\section{Perturbative solution of boson star}
\label{largej}

In this appendix, we perturbatively construct the boson stars for small deformations from pure AdS.
This has been considered in similar contexts \cite{Bizon:2011gg,Dias:2011ss,Horowitz:2014hja,Dias:2016ewl,Martinon:2017uyo,Dias:2017tjg}.
In particular, we consider the large-$j$ limit.

\subsection{Higher order perturbation}

We consider the higher order perturbation around global AdS as
\begin{equation}
 \Phi_j(r)=\Phi_1(r)\epsilon +\Phi_3(r) \epsilon^3+\cdots
\end{equation}
and
\begin{equation}
 \begin{pmatrix}
f(r) \\
g(r) \\
h(r) \\
\beta(r)
\end{pmatrix}
=
\begin{pmatrix}
1 \\
1 \\
\Omega_0 \\
1
\end{pmatrix}
+
\begin{pmatrix}
f_2(r) \\
g_2(r) \\
h_2(r) \\
\beta_2(r)
\end{pmatrix}
\epsilon^2+\cdots\ .
\end{equation}
We set $\alpha(r)=1$ for the boson star.
The fundamental normal mode frequency of the test scalar field in pure AdS is $\Omega_0=1+2/j$.

The equation in the first order in $\epsilon$ is given by
\begin{equation}
 \Phi_1''+\frac{3+5r^2}{r(1+r^2)}\Phi_1'-\frac{4\{j(j+1)-(4+3j)r^2\}}{r^2(1+r^2)^2}\Phi_1=0\ .
\label{phieq}
\end{equation}
This can be solved by
\begin{equation}
 \Phi_1(r)=\phi(r)\equiv \frac{1}{(1+r^2)^2}\left(\frac{r^2}{1+r^2}\right)^j\ ,
 \label{1stsol}
\end{equation}
which is regular at the origin ($r=0$) and infinity ($r=\infty$):
\begin{equation}
 \phi(r)\sim r^{2j}\quad (r\to 0)\ ,\qquad
 \phi(r)\sim \frac{1}{r^4}\quad (r\to \infty)\ .
\end{equation}
The other solution to Eq.~(\ref{phieq}) is
\begin{equation}
 \tilde{\phi}(r)=\phi(r)\int\frac{r}{r^3(1+r^2)\phi^2}\ .
\end{equation}
However, this is singular at the origin and infinity as
\begin{equation}
 \tilde{\phi}(r)\sim r^{-2j-2}\quad (r\to 0)\ ,\qquad
 \tilde{\phi}(r)\sim 1 \quad (r\to \infty)\ .
\end{equation}
Therefore, we adopt $\phi$ (\ref{1stsol}) as the first order solution.
The solution satisfies
\begin{equation}
 \phi'=\frac{2(j-2r^2)}{r(1+r^2)}\phi\ .
\label{dphieq}
\end{equation}
We will use this equation to eliminate $\phi'$ in the following calculations.

The equations in the second order in $\epsilon$ are
\begin{equation}
\begin{split}
 &h_2''+\frac{5}{r}h_2'=\frac{8j(j+2)}{r^2(1+r^2)}\phi^2\ ,\\
&\beta_2''+\frac{3+5r^2}{r(1+r^2)}\beta_2'-\frac{8}{r^2(1+r^2)}\beta_2=-\frac{4j(2j-1)}{r^2(1+r^2)}\phi^2\ ,\\
&g_2''+\frac{2(1+2r^2)}{r(1+r^2)}g_2
+\frac{1}{3(1+r^2)}\beta_2'
+\frac{10}{3r(1+r^2)}\beta_2=
-\frac{8(j+2r^2)}{3r(1+r^2)}\phi^2
\ ,\\
&f_2'+\frac{2(1+2r^2)}{r(1+r^2)}g_2+\frac{2+3r^2}{3(1+r^2)}\beta_2'+\frac{2}{3r(1+r^2)}\beta_2
=-\frac{4(j-4r^2)}{3r(1+r^2)}\phi^2\ ,
\end{split}
\label{2ndeq}
\end{equation}
where, in the right hand side, we have eliminated $\phi'$ by using Eq.~(\ref{dphieq}).
The solutions to them can be decomposed into particular and homogeneous solutions as
\begin{equation}
 \begin{pmatrix}
f_2 \\
g_2 \\
h_2 \\
\beta_2
\end{pmatrix}
= \begin{pmatrix}
f_2^\textrm{p} \\
g_2^\textrm{p}\\
h_2^\textrm{p} \\
\beta_2^\textrm{p}
\end{pmatrix}
+
\begin{pmatrix}
f_2^\textrm{hom} \\
g_2^\textrm{hom} \\
h_2^\textrm{hom} \\
\beta_2^\textrm{hom}
\end{pmatrix}\ ,
\end{equation}
where
\begin{equation}
\begin{split}
&h_2^\textrm{p}=2j(j+1)\left[\int_0^r \frac{\phi^2}{r(1+r^2)}\mathrm{d}r-\frac{1}{r^4}\int_0^r \frac{r^3\phi^2}{1+r^2}\mathrm{d}r\right]\ ,\\
&\beta_2^\textrm{p}=j(2j-1)\left[
\frac{1}{r^4} \int^r_0 r F(r)\phi^2 \mathrm{d}r-F(r) \int^r_0 \frac{\phi^2}{r^3} \mathrm{d}r
\right]\ ,\\
&g_2^\textrm{p}=-\frac{1}{3(1+r^2)}\left[
\beta_2^\textrm{p}+\frac{8}{r^2}\int^r_0\{(j+2r^2)r\phi^2+r\beta_2^\textrm{p}\}\mathrm{d}r
\right]\ ,\\
&f_2^\textrm{p}=-\frac{2+3r^2}{3(1+r^2)}\beta_2^\textrm{p} \\
&\hspace{1cm}-2\int^r_0 \frac{1}{r(1+r^2)}\left\{
(1+2r^2)g_2^\textrm{p}+\frac{2}{3}(j-4r^2)\phi^2+\frac{1}{3(1+r^2)}\beta_2^\textrm{p}
\right\}\mathrm{d}r \ ,
\end{split}
\label{spsols}
\end{equation}
and
\begin{equation}
\begin{split}
&h_2^\textrm{hom}=C_h\ ,\qquad
\beta_2^\textrm{hom}=C_\beta A(r)\ ,\\
&g_2^\textrm{hom}=C_\beta\frac{(4r^2+3)A(r) - 4 r^2}{3 (1+r^2)}\ ,\qquad
f_2^\textrm{hom}=C_f+C_\beta \frac{r^2A(r)}{3(1+r^2)}\ .
\end{split}
\label{homsols}
\end{equation}
Here, we defined
\begin{equation}
 A(r)\equiv 1-\frac{2}{r^2}+\frac{2\ln(1+r^2)}{r^4}\ .
\end{equation}

Since we set the lower bound of the integral at $r=0$,
the regularity at the origin is guaranteed in these expressions.
Integration constants $C_\beta$ and $C_f$ will be chosen so that $\beta_2\to 0$ and $f_2\to 0$ ($r\to\infty$) are satisfied.
$C_h$ will be determined from the third order equation.

The third order equation is
\begin{equation}
\Phi_3''+\frac{3+5r^2}{r(1+r^2)}\Phi_3'-\frac{4\{j(j+1)-(4+3j)r^2\}}{r^2(1+r^2)^2}\Phi_3=S^3\ ,
\end{equation}
where
\begin{multline}
 S^3=\frac{4\phi}{(1+r^2)^2}\bigg[
(j+2)^2f_2
-\frac{4r^4-2r^2-5jr^2+j^2}{r^2}g_2\\
-2j(j+2)h_2-\frac{j^2(1+r^2)+2r^2-j}{r^2}\beta_2+\frac{j(j-2r^2)}{r^2}\phi^2
\bigg]
\ .
\end{multline}
We have eliminated $f_2'$, $g_2'$ and $\phi'$ in the source term by using Eqs.~(\ref{dphieq}) and (\ref{2ndeq}).
The solution in this order is written as
\begin{equation}
 \Phi_3=C_\phi \phi(r)
-\phi(r)\int_0^r r^3(1+r^2)S^3 \tilde{\phi}(r) \mathrm{d}r
+\tilde{\phi}(r)\int_0^r r^3(1+r^2)S^3 \phi(r) \mathrm{d}r \ .
\end{equation}
The first and second terms are regular at the origin and infinity.
The last term is also regular at the origin but it behaves as $\sim 1$ at the infinity.
To satisfy the source free boundary condition at the infinity, therefore, we need
\begin{equation}
 \int_0^\infty r^3(1+r^2)S^3 \phi(r) \mathrm{d}r = 0\ .
\label{Chcond}
\end{equation}
This equation determines $C_h$ as announced.

\subsection{Large-$j$ expansion of the perturbative solution}

The integral expressions of the perturbative results obtained above can be evaluated in the large-$j$ limit in powers of $1/j$.

For the large-$j$ expansion, it is convenient to introduce a new coordinate $x$ as
\begin{equation}
 \frac{r^2}{1+r^2}=\exp(-x^2/j)\ .
\end{equation}
In the new coordinate, $r=0$ and $\infty$ correspond to $x=\infty$ and $0$, respectively.
In terms of $x$, the first order solution (\ref{1stsol}) is rewritten as
\begin{equation}
 \phi=(1-e^{-x^2/j})^2 e^{-x^2}\ .
\end{equation}
Because of the exponential factor $e^{-x^2}$, this function is highly suppressed in $x\gg 1$, i.e.~$r \sim 0$.
Assuming $x\lesssim 1$, we can expand the prefactor $(1-e^{-x^2/j})^2$ by $1/j$ as
\begin{equation}
 \phi=e^{-x^2}\left(\frac{x^4}{j^2}-\frac{x^6}{j^3}+\frac{7x^8}{12j^4}-\frac{x^{10}}{4j^5}+\cdots\right)\ .
\end{equation}

Let us consider the large-$j$ approximation of the second order solution.
For example, the first term in $h_2^\textrm{p}$ (\ref{spsols}) is written as
\begin{equation}
 \int_0^r \frac{\phi^2}{r(1+r^2)}\mathrm{d}r
= \frac{1}{j}\int^\infty_x x (1-e^{-x^2/j})^4 e^{-2x^2} \mathrm{d}x
\end{equation}
The integrand contains an exponential factor $e^{-2x^2}$.
Hence, we assume $x\lesssim 1$ and expand the integrand by $1/j$ as
\begin{equation}
\begin{split}
&\int^\infty_x e^{-2x^2}\left(\frac{x^9}{j^5} - \frac{2x^{11}}{j^6} + \frac{13x^{13}}{6 j^7} - \frac{5x^{15}}{3 j^8} + \cdots\right)\mathrm{d}x\\
=&\, e^{-2x^2}\bigg[\frac{1}{8j^5}(2x^{8}+4x^{6}+6x^{4}+6x^{2}+3)\\
&-\frac{1}{8j^6}(4x^{10}+10x^{8}+20x^{6}+30x^{4}+30x^{2}+15)\\
&+\frac{13}{96j^7}(4x^{12}+12x^{10}+30x^{8}+60x^{6}+90x^{4}+90x^{2}+45)\\
&-\frac{5}{96j^8}(8x^{14}+28x^{12}+84x^{10}+210x^{8}+420x^{6}+630{x}^{4}+630x^{2}+315)\\
&+\cdots\bigg]
\ .
\end{split}
\end{equation}
We can do the same procedure for the other integrals in Eq.~(\ref{spsols}).
Since the integrals always produce the exponential factor $e^{-2x^2}$,
the functions of $r$ outside of the integrals can be expanded by $1/j$
after changing coordinates from $r$ to $x$.
As a result, we have
\begin{equation}
\begin{split}
e^{2x^2}h_2^\textrm{p}
&=\frac{1}{4j^3} (x^{2}+1)( 2x^{4}+3x^{2}+3)\\
&-\frac{1}{8j^4}(6x^{8}+16x^{6}+31x^{4}+36x^{2}+18)\\
&+\frac{1}{48j^5}(30x^{10}+89x^{8}+216x^{6}+393x^{4}+450x^{2}+225)\\
&-\frac{1}{96j^6}(36x^{12}+120x^{10}+343x^{8}+804x^{6}+1425x^{4}+1620x^{2}+810)\\
&+\frac{1}{2880j^7}(516x^{14}+1930x^{12}+6336x^{10}+17805x^{8}\\
&\hspace{3cm}+41040x^{6}+71775x^{4}+81270x^{2}+40635)
+\cdots \ ,
\end{split}
\label{h2expad}
\end{equation}
\begin{equation}
\begin{split}
e^{2x^2}\beta_2^\textrm{p}
&=-\frac{1}{4j^3}\left( x^{2}+1 \right)  \left( 2x^{4}+3x^{2}+3 \right)\\
&+\frac{1}{8j^4}(6x^{8}+24x^{6}+47x^{4}+48x^{2}+18)\\
&-\frac{1}{48j^5}(30x^{10}+161x^{8}+456x^{6}+3(259+24y)x^4+666x^{2}+225)\\
&+\frac{1}{96j^6}(36x^{12}+240x^{10}+887x^{8}+ 12 (181-12y) x^{6}\\
&\hspace{5cm}+ 3(1151+144y) x^{4}+2520x^{2}+810)\\
&-\frac{1}{2880j^7}(516x^{14}+4090x^{12}+18576x^{10}\\
&\hspace{2cm}+ 105(565+24y) x^{8}+ 540(245-24y) x^{6}\\
&\hspace{3cm}+675(299+40y) x^{4}+129870x^{2}+40635)
+\cdots \ ,
\end{split}
\end{equation}
\begin{equation}
\begin{split}
e^{2x^2}g_2^\textrm{p}
&=-\frac{1}{4j^4}x^{2} ( x^4-1)( 2x^{2}+5)\\
&+\frac{1}{8j^5}x^2(8x^{8}+29x^{6}+5x^{4}-(37+16y) x^{2}-30)\\
&-\frac{1}{48j^6}x^2(52x^{10}+243x^{8}+169x^{6}\\
&\hspace{5cm}-(241-72y) x^{4}-96(8+3y) x^{2}-375)\\
&+\frac{1}{92j^7}x^2(80x^{12}+454x^{10}+531x^{8}-8(14-11y) x^{6}\\
&\hspace{3cm}
-(1853-432y) x^{4}-5(743+240y) x^{2}-1350)
+\cdots \ ,
\end{split}
\end{equation}
\begin{equation}
\begin{split}
e^{2x^2}f_2^\textrm{p}&=\frac{1}{4j^3} ( x^{2}+1 )  ( 2x^{4}+3x^{2}+3 ) \\
&-\frac{1}{8j^4}(10x^{8}+18x^{6}+27x^{4}+30x^{2}+18)\\
&+\frac{1}{48j^5}(78x^{10}+95x^{8}+150x^{6}+ 3(81-8y) x^{4}+342x^{2}+225)\\
&-\frac{1}{96j^6}(140x^{12}+102x^{10}+105x^{8}\\
&\hspace{4cm}+186x^{6}+ 3(191-48y) x^{4}+1170x^{2}+810)\\
&+\frac{1}{2880j^7}(2916x^{14}+910x^{12}-2214x^{10}- 15(549+8y) x^{8}\\
&\hspace{1cm}-7530x^{6}+ 75(221-120y) x^{4}+56970x^{2}+40635)
+\cdots \ ,
\end{split}
\label{f2expad}
\end{equation}
where we defined
\begin{equation}
 y=e^{2x^{2}}\textrm{Ei}_1(2x^{2})\ ,\qquad
\textrm{Ei}_1(z)=\int^\infty_1 \frac{e^{-tz}}{t}\mathrm{d}t
\ .
\end{equation}
At the origin and infinity, this behaves as
\begin{equation}
 y\simeq -\gamma - \ln 2  - 2\ln x \quad(x\to 0)\ ,\qquad
 y\simeq \frac{1}{2x^2}\quad (x\to \infty)\ ,
\end{equation}
where $\gamma$ is Euler's constant.
Near the infinity ($x=0$), the second order solutions approach
\begin{align}
&h_2^\textrm{p}\to \frac{3}{4j^3}-\frac{9}{4j^4}+\frac{75}{16j^5}-\frac{135}{16j^6}+\frac{903}{64j^7}+\cdots\ ,\\
&\beta_2^\textrm{p}\to -\frac{3}{4j^3}+\frac{9}{4j^4}-\frac{75}{16j^5}+\frac{135}{16j^6}-\frac{903}{64j^7}+\cdots\ ,\\
&g_2^\textrm{p}\to 0\ ,\\
&f_2^\textrm{p}\to \frac{3}{4j^3}-\frac{9}{4j^4}+\frac{75}{16j^5}-\frac{135}{16j^6}+\frac{903}{64j^7}+\cdots .
\end{align}
Therefore, from the boundary condition for the second order perturbation, $\beta_2\to 0$ and $f_2\to 0$ at $r\to\infty$, we can determine the integration constants $C_\beta$ and $C_f$ as
\begin{equation}
\begin{split}
&C_\beta=\frac{3}{4j^3}-\frac{9}{4j^4}+\frac{75}{16j^5}-\frac{135}{16j^6}+\frac{903}{64j^7}+\cdots\ ,\\
&C_f=-\frac{4}{3}\left(\frac{3}{4j^3}-\frac{9}{4j^4}+\frac{75}{16j^5}-\frac{135}{16j^6}+\frac{903}{64j^7}+\cdots\right)\ .
\end{split}
\end{equation}
To determine $C_h$, we substitute the particular solutions~(\ref{h2expad}-\ref{f2expad}) and homogeneous solutions~(\ref{homsols}) into Eq.~(\ref{Chcond}).
Then, expanding the integrand by $1/j$ after moving to the $x$-coordinate, we can carry out the integration.
As the result, we obtain $C_h$ as
\begin{multline}
 C_h=-\frac{3}{4j^3}+\frac{9}{4j^4}- \frac{3}{2j^5}
\left(\frac{29}{16}+\ln(e^\gamma j)\right)\\
+\frac {21}{4j^6} \left(-\frac{137}{448}+\ln(e^\gamma j)\right)\\
-\frac{12}{j^7} \left(-\frac{16079}{12288}+\ln(e^\gamma j)\right)
+\cdots\ .
\end{multline}

Near the infinity, the second order solutions are expanded as
\begin{equation}
 f_2=\frac{c_f}{r^4}+\cdots\ ,\quad
 g_2=\frac{c_g}{r^4}+\cdots\ ,\quad
 h_2=\Omega_2+\frac{c_h}{r^4}+\cdots\ ,\quad
\beta=\frac{c_\beta}{r^4}+\cdots\ ,
\end{equation}
where
\begin{equation}
\begin{split}
 c_f=&-\frac{1}{4j}-\frac{3}{8j^2}+\frac{1}{2j^3}\left(\frac{9}{8}+\ln(2e^\gamma j)\right)\\
&-\frac{3}{2j^4}\left(-\frac{13}{48}+\ln(2e^\gamma j)\right)
+\frac{25}{8j^5}\left(-\frac{623}{600}+\ln(2e^\gamma j)\right)+\cdots\ ,\\
c_g=&-\frac{2}{j^2}+\frac{2}{j^3}(1+\ln(2e^\gamma j))\\
&\qquad -\frac{6}{j^4}\left(-\frac{1}{2}+\ln(2e^\gamma j)\right)
+\frac{25}{2j^5}\left(-\frac{77}{60}+\ln(2e^\gamma j)\right)+\cdots\ ,\\
c_h=&-\frac{1}{4j}+\frac{5}{8j^2}-\frac{19}{16j^3}+\frac{65}{32j^4}-\frac{211}{64j^5}+\cdots\ ,\\
c_\beta=&
\frac{1}{4j}
-\frac{13}{8j^2}
+\frac{3}{2j^3}\left(\frac{23}{24}+\ln(2e^\gamma j)\right)\\
&\quad -\frac{9}{2j^4}\left(-\frac{83}{144}+\ln(2e^\gamma j)\right)
+\frac{75}{8j^5}\left(-\frac{273}{200}+\ln(2e^\gamma j)\right)
+\cdots\ .
\end{split}
\end{equation}
The second order contribution of the normal mode frequency is
\begin{multline}
 \Omega_2=
-\frac{3}{2j^5}\left(-\frac{21}{16} + \ln(e^\gamma j)\right)
+\frac{21}{4j^6}\left(-\frac{857}{448} +\ln(e^\gamma j)\right)\\
-\frac{12}{j^7}\left(-\frac{30527}{12288} +\ln(e^\gamma j)\right)+\cdots\ .
\end{multline}
From $c_f,c_\beta,c_h$, we can compute the second order contribution of the mass $E_2$ and angular momentum $J_2$ of the boson star as
\begin{align}
&\frac{8G}{\pi}E_2=c_\beta-3c_f=\frac{1}{j}-\frac{1}{2j^2}-\frac{1}{4j^3}+\frac{11}{8j^4}-\frac{49}{16j^5}+\cdots\ ,\\
&\frac{8G}{\pi}J_2=-4c_h=\frac{1}{j}-\frac{5}{2j^2}+\frac{19}{4j^3}-\frac{65}{8j^4}+\frac{211}{16j^5}+\cdots\ .
\end{align}
One can check that these satisfy the first law at all orders in the $1/j$-expansion:
\begin{equation}
 E_2=\Omega_0 J_2\ .
\end{equation}

\bibliography{bib_multiplet}

\end{document}